\documentclass[a4paper,11pt]{article}
\usepackage{jheppub}
\usepackage[utf8]{inputenc}

\usepackage{amsfonts}
\usepackage{amssymb}
\usepackage{mathrsfs}
\usepackage{mathtools}

\usepackage{tikz}
\usetikzlibrary{
  decorations.markings,
  decorations.pathmorphing,
}

\def\graphwidth{2.0}
\def\graphheight{1.2}
\def\cutval{0.7}

\def\contourgap{0.12}
\def\axescolor{black}
\def\contourcolor{red}
\def\surfacecolor{blue!30}
\def\innercolor{blue!10}

\tikzset{cutstyle/.style={decorate, decoration={zigzag, segment length=6, amplitude=2}, draw=black}}
\tikzset{arrow data/.style 2 args={decoration={markings, mark=at position #1 with \arrow{#2}}, postaction=decorate}}
\tikzset{partial ellipse/.style args={#1:#2:#3}{insert path={+ (#1:#3) arc (#1:#2:#3)}}}

\tikzset{pics/torus/.style n args={3}{
    code = {
    \providecolor{pgffillcolor}{rgb}{1,1,1}
    \begin{scope}[
        yscale=cos(#3),
        outer torus/.style = {draw,line width/.expanded={\the\dimexpr2\pgflinewidth+#2*2},line join=round},
        inner torus/.style = {draw=pgffillcolor,line width={#2*2}}
    ]
    \draw[outer torus] circle(#1);\draw[inner torus] circle(#1);
    \draw[outer torus] (190:#1) arc (190:350:#1);\draw[inner torus,line cap=round] (180:#1) arc (180:360:#1);
    \end{scope}
    }
}}

\newcommand\drawaxes[2]{
    \draw[-latex, \axescolor] (-\graphwidth, 0) -- (+\graphwidth, 0) node[below left] {#1};
    \draw[-latex, \axescolor] (0, -\graphheight) -- (0, +\graphheight) node[below left] {#2};
}
\newcommand\drawxmark[3]{
   \draw[fill] (#1, 0) circle (.7pt)
    node[black, shift=#2] {#3};
}

\newcommand\dogbone{
    \begin{tikzpicture}[scale=2, thick]
    \drawaxes{$\operatorname{Re}\,s$}{$\operatorname{Im}\,s$}
    \drawxmark{\cutval}{(-40:0.7)}{$t^{-\frac12}$}
    \drawxmark{-\cutval}{(130:0.7)}{$-t^{-\frac12}$}
    \draw[cutstyle, thick] (-\cutval, 0) -- (+\cutval, 0);
    \draw[\contourcolor, ultra thick, arrow data={0.3}{latex}, arrow data={0.8}{latex}]
        (\cutval,-\contourgap) -- (-\cutval,-\contourgap) arc (270:90:\contourgap) -- (\cutval,\contourgap) node[above right] {$\mathfrak{S}$} arc (90:-90:\contourgap);
    \end{tikzpicture}
}

\newcommand\hankel{
    \begin{tikzpicture}[scale=2, thick]
    \draw[-latex, \axescolor] (-\graphwidth/2, 0) -- (+\graphwidth, 0) node[right] {$\operatorname{Re}x$};
    \draw[-latex, \axescolor] (0, -\graphheight/1.2) -- (0, +\graphheight/1.2) node[below left] {$\operatorname{Im}x$};
    \drawxmark{\cutval}{(-130:0.5)}{$1$};
    \draw[cutstyle] (+\cutval, 0) -- ({+\graphwidth-0.1}, 0);
    \draw[\contourcolor, ultra thick, arrow data={0.25}{latex}, arrow data={0.8}{latex}]
        (\graphwidth,-\contourgap) -- (\cutval,-\contourgap) arc (270:90:\contourgap) -- (\graphwidth,\contourgap);
    \end{tikzpicture}
}

\newcommand\lateralborel{
    \begin{tikzpicture}[scale=2, thick]
    \draw[-latex, \axescolor] (-\graphwidth/2, 0) -- (+\graphwidth, 0) node[right] {$\operatorname{Re}\zeta$};
    \draw[-latex, \axescolor] (0, -\graphheight/1.2) -- (0, +\graphheight/1.2) node[below left] {$\operatorname{Im}\zeta$};
    \drawxmark{\cutval}{(-120:0.6)}{$2/u$};
    \draw[cutstyle] (+\cutval, 0) -- ({+\graphwidth-0.1}, 0);
    \draw[\contourcolor, ultra thick, arrow data={0.3}{latex}, arrow data={0.8}{latex}]
        (0,0) -- (\graphwidth,-{1.5*\contourgap}) node[shift=(-120:0.6)] {$\mathcal{S}_{0^-}$}
        (0,0) -- (\graphwidth,+{1.5*\contourgap}) node[shift=(+120:0.6)] {$\mathcal{S}_{0^+}$};
    \end{tikzpicture}
}

\newcommand\disk{
    \begin{tikzpicture}[scale=2, rotate=-90, thick]
    \draw[black!50, dashed] (-1,-0.02) to[in=127, out=-50] (-0.71,-0.35);
    \draw[black!50, dashed] (1,-0.02) to[in=53, out=-130] (0.71,-0.35);
    \draw[black, fill=\surfacecolor]
    (0.71,-0.35) to[in=73, out=-127] (0.265,-1.2)
    to[out=-107, in=0] (0,-1.45)
    to[out=180, in=-73] (-0.265,-1.2)
    to[out=107, in=-50] (-0.71,-0.35) -- cycle;
    \draw[black, fill=\innercolor] plot [smooth cycle, tension=0.8] coordinates{
        (1/1.4,-0.33) (0.68,-0.38) (0.62,-0.42) (0.5,-0.46) (0.36,-0.44) (0.22,-0.52)
        (0.07,-0.47) (-0.07,-0.53) (-0.22,-0.49) (-0.36,-0.54) (-0.5,-0.44) (-0.62,-0.42)
        (-0.68,-0.38) (-1/1.4,-0.33) (-0.68,-0.28) (-0.62,-0.26) (-0.5,-0.18) (-0.36,-0.2)
        (-0.22,-0.12) (-0.07,-0.16) (0.07,-0.13) (0.22,-0.16) (0.36,-0.11) (0.5,-0.23)
        (0.62,-0.22) (0.68,-0.28)
    };
    \draw (0,0.05) node {$\beta/\epsilon$};
    \draw[black!50, fill=none, dashed] (0,0) ellipse (1 and 0.25);
    \end{tikzpicture}
}

\newcommand\trumpet{
    \begin{tikzpicture}[scale=2, rotate=-90, thick]
    \draw[thin, \surfacecolor, fill=\surfacecolor]
    (0.71,-0.35) to[in=80, out=-127] (1/5,-1.5)
    to[bend left=26] (-1/5,-1.5)
    to[in=-50, out=100] (-0.71,-0.35);
    \draw[black!50, dashed] (1,-0.02) to[in=53, out=-130] (0.71,-0.35);
    \draw[black] (0.71,-0.35) to[in=80, out=-127] (1/5,-1.5);
    \draw[black!50, dashed] (-1,-0.02) to[in=127, out=-50] (-0.71,-0.35);
    \draw[black] (-0.71,-0.35) to[in=100, out=-53] (-1/5,-1.5);
    \draw[black, fill=\innercolor] plot [smooth cycle, tension=0.8] coordinates{
        (1/1.4,-0.33) (0.68,-0.38) (0.62,-0.42) (0.5,-0.46) (0.36,-0.44) (0.22,-0.52)
        (0.07,-0.47) (-0.07,-0.53) (-0.22,-0.49) (-0.36,-0.54) (-0.5,-0.44) (-0.62,-0.42)
        (-0.68,-0.38) (-1/1.4,-0.33) (-0.68,-0.28) (-0.62,-0.26) (-0.5,-0.18) (-0.36,-0.2)
        (-0.22,-0.12) (-0.07,-0.16) (0.07,-0.13) (0.22,-0.16) (0.36,-0.11) (0.5,-0.23)
        (0.62,-0.22) (0.68,-0.28)
    };
    \draw (-0.03,-1.65) node {$b$};
    \draw (0,0.05) node {$\beta/\epsilon$};
    \draw[black!50, fill=none, dashed] (0,0) ellipse (1 and 0.25);
    \draw[black, dashed] (0,-1.5) [partial ellipse=0:180:1/5 and 0.25/5];
    \draw[black] (0,-1.5) [partial ellipse=-180:0:1/5 and 0.25/5];
    \end{tikzpicture}
}

\newcommand\cylinder{
    \begin{tikzpicture}[scale=2, rotate=-90, thick]
    \draw[thin, \surfacecolor, fill=\surfacecolor]
    (0.71,-0.35) to[in=80, out=-127] (1/5,-1.5)
    to[bend left=26] (-1/5,-1.5)
    to[in=-50, out=100] (-0.71,-0.35);
    \draw[black!50, dashed] (1,-0.02) to[in=53, out=-130] (0.71,-0.35);
    \draw[black] (0.71,-0.35) to[in=80, out=-127] (1/5,-1.5);
    \draw[black!50, dashed] (-1,-0.02) to[in=127, out=-50] (-0.71,-0.35);
    \draw[black] (-0.71,-0.35) to[in=100, out=-53] (-1/5,-1.5);
    \draw[black, fill=\innercolor] plot [smooth cycle, tension=0.8] coordinates{
        (1/1.4,-0.33) (0.68,-0.38) (0.62,-0.42) (0.5,-0.46) (0.36,-0.44) (0.22,-0.52)
        (0.07,-0.47) (-0.07,-0.53) (-0.22,-0.49) (-0.36,-0.54) (-0.5,-0.44) (-0.62,-0.42)
        (-0.68,-0.38) (-1/1.4,-0.33) (-0.68,-0.28) (-0.62,-0.26) (-0.5,-0.18) (-0.36,-0.2)
        (-0.22,-0.12) (-0.07,-0.16) (0.07,-0.13) (0.22,-0.16) (0.36,-0.11) (0.5,-0.23)
        (0.62,-0.22) (0.68,-0.28)
    };
    \draw (0,0) node {$u_1$};
    \draw (0.4,-1.6) node {$b$};
    \draw[black!50, fill=none, dashed] (0,0) ellipse (1 and 0.25);
    \draw[black, dashed] (0,-1.5) [partial ellipse=0:180:1/5 and 0.25/5];
    \draw[black] (0,-1.5) [partial ellipse=-180:0:1/5 and 0.25/5];
    \begin{scope}[yshift=-90, rotate=180]
        \draw[thin, \surfacecolor, fill=\surfacecolor]
        (0.71,-0.35) to[in=80, out=-127] (1/5,-1.5)
        to[bend left=26] (-1/5,-1.5)
        to[in=-50, out=100] (-0.71,-0.35);
        \draw[black!50, dashed] (1,-0.02) to[in=53, out=-130] (0.71,-0.35);
        \draw[black] (0.71,-0.35) to[in=80, out=-127] (1/5,-1.5);
        \draw[black!50, dashed] (-1,-0.02) to[in=127, out=-50] (-0.71,-0.35);
        \draw[black] (-0.71,-0.35) to[in=100, out=-53] (-1/5,-1.5);
        \draw[black!50, fill=none, dashed] (0,0) ellipse (1 and 0.25);
        \draw[black, fill=\surfacecolor] plot [smooth, tension=0.8] coordinates{
            (-0.68,-0.38) (-1/1.4,-0.33) (-0.68,-0.28) (-0.62,-0.26) (-0.5,-0.18) (-0.36,-0.2)
            (-0.22,-0.12) (-0.07,-0.16) (0.07,-0.13) (0.22,-0.16) (0.36,-0.11) (0.5,-0.23)
            (0.62,-0.22) (0.68,-0.28) (0.7,-0.30) (1/1.4,-0.33) (0.68,-0.38)
        };
        \draw[black, fill=none, dashed] plot [smooth, tension=0.8] coordinates {
            (1/1.4,-0.33) (0.68,-0.38) (0.62,-0.42) (0.5,-0.46) (0.36,-0.44) (0.22,-0.52)
            (0.07,-0.47) (-0.07,-0.53) (-0.22,-0.49) (-0.36,-0.54) (-0.5,-0.44) (-0.62,-0.42)
            (-0.68,-0.38) (-1/1.4,-0.33)
        };
        \draw (0,0) node {$u_2$};
        \draw[black!50, dotted] (0,0) [partial ellipse=-122:-58:1 and 0.25];
        \draw[black, fill=\innercolor] (0,-1.5) ellipse (1/5 and 0.25/5);
    \end{scope}
    \end{tikzpicture}
}

\newcommand\genusone{
    \begin{tikzpicture}[scale=2, rotate=-90, thick]
    \draw[thin, \surfacecolor, fill=\surfacecolor]
    (0.71,-0.35) to[in=80, out=-127] (1/5,-1.5)
    to[bend left=26] (-1/5,-1.5)
    to[in=-50, out=100] (-0.71,-0.35);
    \draw[black!50, dashed] (1,-0.02) to[in=53, out=-130] (0.71,-0.35);
    \draw[black] (0.71,-0.35) to[in=80, out=-127] (1/5,-1.5);
    \draw[black!50, dashed] (-1,-0.02) to[in=127, out=-50] (-0.71,-0.35);
    \draw[black] (-0.71,-0.35) to[in=100, out=-53] (-1/5,-1.5);
    \draw[black, fill=\innercolor] plot [smooth cycle, tension=0.8] coordinates{
        (1/1.4,-0.33) (0.68,-0.38) (0.62,-0.42) (0.5,-0.46) (0.36,-0.44) (0.22,-0.52)
        (0.07,-0.47) (-0.07,-0.53) (-0.22,-0.49) (-0.36,-0.54) (-0.5,-0.44) (-0.62,-0.42)
        (-0.68,-0.38) (-1/1.4,-0.33) (-0.68,-0.28) (-0.62,-0.26) (-0.5,-0.18) (-0.36,-0.2)
        (-0.22,-0.12) (-0.07,-0.16) (0.07,-0.13) (0.22,-0.16) (0.36,-0.11) (0.5,-0.23)
        (0.62,-0.22) (0.68,-0.28)
    };
    \draw (0,0) node {$u$};
    \draw (0.4,-1.6) node {$b$};
    \draw[black!50, fill=none, dashed] (0,0) ellipse (1 and 0.25);
    \draw[black, dashed] (0,-1.5) [partial ellipse=0:180:1/5 and 0.25/5];
    \draw[black] (0,-1.5) [partial ellipse=-180:0:1/5 and 0.25/5];
    \begin{scope}[yshift=-90, rotate=180]
        \pic[fill=\surfacecolor,draw=black, xshift=20]{torus={1.3cm}{4mm}{65}};
        \draw[\surfacecolor, fill=\surfacecolor]
        (0.40,-0.84) to[in=80, out=-115] (1/5,-1.5)
        to (-1/5,-1.5) to[in=-70, out=100] (-0.40,-0.85) --cycle;
        \draw
        (0.409,-0.83) to[in=80, out=-112] (1/5,-1.5)
        (-1/5,-1.5) to[in=-70, out=100] (-0.405,-0.84);
        \draw[black, fill=\innercolor] (0,-1.5) ellipse (1/5 and 0.25/5);
        \draw (0.4,0.6) node {$V_{1,1}$};
        \end{scope}
    \end{tikzpicture}
}

\newcommand\sffdisk{
    \begin{tikzpicture}[yscale=1, xscale=1, rotate=-90, thick]
    \draw[black, fill=\surfacecolor]
    (0.71,-0.35) to[in=73, out=-127] (0.265,-1.0)
    to[out=-107, in=0] (0,-1.25)
    to[out=180, in=-73] (-0.265,-1.0)
    to[out=107, in=-50] (-0.71,-0.35) -- cycle;
    \draw[black, fill=\innercolor] plot [smooth cycle, tension=0.8] coordinates
    {(1/1.4,-0.33) (0.68,-0.38) (0.62,-0.42) (0.5,-0.46) (0.36,-0.44) (0.22,-0.52)
    (0.07,-0.47) (-0.07,-0.53) (-0.22,-0.49) (-0.36,-0.54) (-0.5,-0.44) (-0.62,-0.42)
    (-0.68,-0.38) (-1/1.4,-0.33) (-0.68,-0.28) (-0.62,-0.26) (-0.5,-0.18) (-0.36,-0.2)
    (-0.22,-0.12) (-0.07,-0.16) (0.07,-0.13) (0.22,-0.16) (0.36,-0.11) (0.5,-0.23)
    (0.62,-0.22) (0.68,-0.28)};
    \begin{scope}[yshift=-85, rotate=180]
    \draw[black, fill=\surfacecolor]
    (0.71,-0.35) to[in=73, out=-127] (0.265,-1.0)
    to[out=-107, in=0] (0,-1.25)
    to[out=180, in=-73] (-0.265,-1.0)
    to[out=107, in=-50] (-0.71,-0.35) -- cycle;
    \draw[black, fill=\surfacecolor] plot [smooth, tension=0.8] coordinates
    {(-0.68,-0.38) (-1/1.4,-0.33) (-0.68,-0.28) (-0.62,-0.26) (-0.5,-0.18) (-0.36,-0.2)
    (-0.22,-0.12) (-0.07,-0.16) (0.07,-0.13) (0.22,-0.16) (0.36,-0.11) (0.5,-0.23)
    (0.62,-0.22) (0.68,-0.28) (0.7,-0.30) (1/1.4,-0.33) (0.68,-0.38)};
    \draw[black, fill=none, dotted] plot [smooth, tension=0.8] coordinates
    {(1/1.4,-0.33) (0.68,-0.38) (0.62,-0.42) (0.5,-0.46) (0.36,-0.44) (0.22,-0.52)
    (.07,-0.47) (-0.07,-0.53) (-0.22,-0.49) (-0.36,-0.54) (-0.5,-0.44) (-0.62,-0.42)
    (-0.68,-0.38) (-1/1.4,-0.33)};
    \end{scope}
    \end{tikzpicture}
}

\newcommand\sffcylinder{
    \begin{tikzpicture}[yscale=1, xscale=1, rotate=-90, thick]
    \draw[thin, \surfacecolor, fill=\surfacecolor]
    (0.71,-0.35) to[in=90, out=-127] (1/5,-1.5) to[out=-90, in=127] (0.71,-3+0.35) to (-0.71,-3+0.35) to[out=53, in=-90] (-1/5,-1.5) to[out=90, in=-53] (-0.71,-0.35);
    \draw[black]
    (0.71,-0.35) to[in=90, out=-127] (1/5,-1.5) to[out=-90, in=127] (0.71,-3+0.35)
    (-0.71,-3+0.35) to[out=53, in=-90] (-1/5,-1.5) to[out=90, in=-53] (-0.71,-0.35);
    \draw[black, fill=\innercolor] plot [smooth cycle, tension=0.8] coordinates
    {(1/1.4,-0.33) (0.68,-0.38) (0.62,-0.42) (0.5,-0.46) (0.36,-0.44) (0.22,-0.52)
    (0.07,-0.47) (-0.07,-0.53) (-0.22,-0.49) (-0.36,-0.54) (-0.5,-0.44) (-0.62,-0.42)
    (-0.68,-0.38) (-1/1.4,-0.33) (-0.68,-0.28) (-0.62,-0.26) (-0.5,-0.18) (-0.36,-0.2)
    (-0.22,-0.12) (-0.07,-0.16) (0.07,-0.13) (0.22,-0.16) (0.36,-0.11) (0.5,-0.23)
    (0.62,-0.22) (0.68,-0.28)};
    \begin{scope}[yshift=-85, rotate=180]
    \draw[black, fill=\surfacecolor] plot [smooth, tension=0.8] coordinates
    {(-0.68,-0.38) (-1/1.4,-0.33) (-0.68,-0.28) (-0.62,-0.26) (-0.5,-0.18) (-0.36,-0.2)
    (-0.22,-0.12) (-0.07,-0.16) (0.07,-0.13) (0.22,-0.16) (0.36,-0.11) (0.5,-0.23)
    (0.62,-0.22) (0.68,-0.28) (0.7,-0.30) (1/1.4,-0.33) (0.68,-0.38)};
    \draw[black, fill=none, dotted] plot [smooth, tension=0.8] coordinates
    {(1/1.4,-0.33) (0.68,-0.38) (0.62,-0.42) (0.5,-0.46) (0.36,-0.44) (0.22,-0.52)
    (.07,-0.47) (-0.07,-0.53) (-0.22,-0.49) (-0.36,-0.54) (-0.5,-0.44) (-0.62,-0.42)
    (-0.68,-0.38) (-1/1.4,-0.33)};
    \end{scope}
    \end{tikzpicture}
}

\preprint{UUITP-25/21}

\title{\huge\boldmath Nonperturbative effects and resurgence\\in JT gravity at finite cutoff}

\author[a]{Luca Griguolo,}
\author[b]{Rodolfo Panerai,}
\author[a]{Jacopo Papalini,}
\author[c]{and Domenico Seminara}

\affiliation[a]{Dipartimento SMFI, Universit\`a di Parma and INFN Gruppo Collegato di Parma, Viale G.P. Usberti 7/A, 43100 Parma, Italy}
\affiliation[b]{Department of Physics and Astronomy, Uppsala University, Box 516, SE-75120 Uppsala, Sweden}
\affiliation[c]{Dipartimento di Fisica, Universit\`a di Firenze and INFN Sezione di Firenze, via G. Sansone 1, 50019 Sesto Fiorentino, Italy} 

\emailAdd{luca.griguolo@unipr.it}
\emailAdd{rodolfo.panerai@physics.uu.se}
\emailAdd{jacopo.papalini@unipr.it}
\emailAdd{seminara@fi.infn.it}

\abstract{We investigate the nonperturbative structure of Jackiw--Teitelboim gravity at finite cutoff, as given by its proposed formulation in terms of a $T\bar{T}$-deformed Schwarzian quantum mechanics. Our starting point is a careful computation of the disk partition function to all orders in the perturbative expansion in the cutoff parameter. We show that the perturbative series is asymptotic and that it admits a precise completion exploiting the analytical properties of its Borel transform, as prescribed by resurgence theory. The final result is then naturally interpreted in terms of the nonperturbative branch of the $T\bar{T}$-deformed spectrum. The finite-cutoff trumpet partition function is computed by applying the same strategy. In the second part of the paper, we propose an extension of this formalism to arbitrary topologies, using the basic gluing rules of the undeformed case. The Weil--Petersson integrations can be safely performed due to the nonperturbative corrections and give results that are compatible with the flow equation associated with the $T\bar{T}$ deformation. We derive exact expressions for general topologies and show that these are captured by a suitable deformation of the Eynard--Orantin topological recursion. Finally, we study the ``slope'' and ``ramp'' regimes of the spectral form factor as functions of the cutoff parameter.}

\begin{document} 
\maketitle
\flushbottom

\section{Introduction}
A crucial problem in quantum gravity is the precise definition of physical observables. Asymptotic quantities, such as the S-matrix in Minkowski space or boundary correlators in AdS, are pretty well understood. Still, it seems essential to extend the set of computable quantities, especially when black holes and cosmological aspects are involved. A direct attempt to obtain more ``local'' observables would consist of defining quantum gravity in a box, imposing suitable boundary conditions on the metric at some finite spatial extent. In the AdS/CFT correspondence, this program should imply the extension of the holographic dictionary to gravitational theories defined on bounded regions of spacetime. 

Unexpectedly, the renewed interest in a class of solvable irrelevant deformations of two-dimensional CFTs, known as the $T\bar{T}$ deformations \cite{Zamolodchikov:2004ce,Smirnov:2016lqw,Cavaglia:2016oda}, has suggested an innovative strategy to address the above issue, at least in low dimensions. It was proposed in \cite{McGough:2016lol} (see also \cite{Kraus:2018xrn}) that $T\bar{T}$-deformed CFTs could realize the holographic dual of AdS$_3$ gravities on a finite patch.  The crucial element in favor of this conjectured duality is that the conformal Ward identity of the relevant CFT translates, in the presence of the deformation, into a second-order functional differential equation that closely resembles the Wheeler--DeWitt equation of AdS$_3$ gravity. This observation suggests a possible identification between the wave functionals of gravitational theories in $d+1$ dimensions and partition functions of $d$-dimensional nontrivially-deformed QFTs. This correspondence has been checked in various ways \cite{Hartman:2018tkw,Gorbenko:2018oov,Guica:2019nzm}, but the consistency of the proposal is still under scrutiny. An unsatisfying feature of this duality is that for large enough energies, the spectrum of the deformed boundary theory becomes complex, implying a potential breakdown of unitarity for the bulk theory. A different and maybe related problem is to better understand the flow induced by the $T\bar{T}$ deformation from the holographic point of view, as coming from integrating out a portion of the asymptotic geometry.

While there have been attempts to generalize the conjecture to higher dimensions \cite{Taylor:2018xcy}, a simpler context, where we can accurately study the status of the proposal, is to consider the finite cutoff version of Jackiw--Teitelboim (JT) gravity \cite{Jackiw:1984je,Teitelboim:1983ux}. In its modern formulation \cite{Maldacena:2016upp}, the JT path integral itself is defined as a limit procedure from a cut-off theory: after imposing Dirichlet boundary conditions at some finite distance, the proper boundary length and the boundary value for the dilaton are scaled appropriately in the large area limit to preserve the boundary degrees of freedom. In so doing, JT gravity reduces to a solvable one-dimensional theory, the Schwarzian quantum mechanics \cite{Bagrets:2016cdf,Stanford:2017thb,Mertens:2017mtv}. The natural expectation is that at finite cutoff the relevant dual formulation is provided by a $T\bar{T}$-deformed version of such a theory \cite{Gross:2019ach,Gross:2019uxi}.

The partition function of JT gravity restricted on a finite AdS$_2$ subregion has been computed in \cite{Iliesiu:2020zld}, using two different approaches based on either canonical or path-integral quantization\footnote{See also the interesting alternative investigation \cite{Stanford:2020qhm}, relying on a completely different method}. The results of both methods are mutually consistent and are directly related to the $T\bar{T}$ deformation of the Schwarzian theory. An important issue addressed in \cite{Iliesiu:2020zld} concerns the spectrum of the deformed theory, which complexifies above a certain energy threshold. As a consequence, the na\"ive integration prescription does not generate a well-defined partition function. A consistent partition function can be obtained instead by adding contributions originating from a nonperturbative branch, but the related spectral density becomes not positive definite, calling for a physical interpretation. Moreover, in the analysis of \cite{Iliesiu:2020zld}, the nonperturbative completion seems to be accompanied by certain ambiguities that the authors cannot wholly fix in their approach. Last but not least, the construction of partition functions for arbitrary topologies configurations, relevant for the nonperturbative definition of JT gravity itself \cite{Saad:2019lba,Stanford:2019vob}, is left unexplored.

In this paper, we reexamine JT gravity at finite cutoff, starting from its definition in terms of the $T\bar{T}$-deformed Schwarzian quantum mechanics. We begin by studying the $T\bar{T}$ flow purely at a perturbative level and compute the entire perturbative series associated with the deformation parameter, both for the disk and the trumpet partition functions. We find that the resulting series has a vanishing radius of convergence and, as such, requires an appropriate nonperturbative completion. We then exploit the standard resurgence technique \cite{Dorigoni:2014hea,Aniceto:2018bis}, using the properties of the lateral Borel resummation, to take into account nonperturbative contributions. This procedure unambiguously brings into the game the nonperturbative configurations associated with the new energy branch and prescribes the correct integration contour. For the disk topology, we obtain the partition function in terms of a modified Bessel function of the first type, an expression already considered in \cite{Iliesiu:2020zld}. The energy spectrum naturally spans a finite interval; however, the associated spectral density is not positive definite. The trumpet partition function experiences an even more dramatic modification: the nonperturbative corrections completely smooth out na\"ive singularity associated with the fact that the cutoff boundary could overlap with the geodesic boundary.

Relying on this observation, in the second part of the paper, we explore the construction of the deformed version of the partition functions for arbitrary topologies, using the same gluing procedure derived for the undeformed theory \cite{Saad:2019lba}. We remark that without the nonperturbative corrections, the relevant gluing integral would be ill-defined. The gluing procedure results in a consistent deformation of the standard Eynard--Orantin recursion relations \cite{Eynard:2007fi} associated with the original theory: the deformed spectral curve and the higher-genus correlation functions are fully compatible with the flow equation of the $T\bar{T}$ deformation, and we find a precise mapping that encodes the flow. We stress that the non-positivity of the input spectral density does not spoil the consistency of the recursion relations, although its actual physical interpretation is still missing in our case. An essential step in our construction is the explicit evaluation of the cylinder partition function: it is closely related to the kernel necessary to engineer the Eynard--Orantin topological recursion formula \cite{Eynard:2007fi} and is responsible for the ``ramp'' growth in the spectral form factor \cite{Saad:2018bqo,Saad:2019lba,Saad:2019pqd}. We derive in this last perspective its late-time behavior and observe the transition between the slope and the ramp phase at finite cutoff. Quite interestingly, the change of regime does not seem to depend on the value of the finite cutoff.

The paper is structured as follows. We start in Section~\ref{SEC:review} by briefly reviewing some generalities on JT gravity and its $T\bar{T}$ deformation. Subsequently, in Section~\ref{SEC:disk_trumpet_from_resurgence}, we compute the perturbative series arising from the deformation and its completion using the theory of resurgence, both for the disk and the trumpet. Section~\ref{SEC:spectrum} is devoted to the spectral properties of the deformed theory: the relevant partition functions are seen arising from a compact spectrum and computed with a suitable integration contour. The spectral density is derived and found to be not positive defined. In Section~\ref{SEC:other_topologies} we construct the partition functions for arbitrary topologies, exploiting the gluing prescription of the undeformed theory. The consistency of this construction with the $T\bar{T}$ flow equation is discussed in Section~\ref{SEC:flow_equation}. The extension of the Eynard--Orantin recursion relations in the deformed case is presented in Section~\ref{SEC:topological_recursion}, opening the possibility to interpret holographically the theory at finite cutoff. Finally, in Section~\ref{SEC:spectral_form_factor}, we study the deformed spectral form factor, deriving the behavior of the ``slope'' and ``ramp'' regimes. Section~\ref{SEC:conclusions} presents our conclusion and the possible extensions of this work. A couple of technical appendices complete the manuscript.

\section{Basics of JT gravity and its \texorpdfstring{$T\bar{T}$}{TT} deformation}\label{SEC:review}
JT gravity is a two-dimensional theory of gravity whose dynamical fields are the metric and a dilaton field $\phi$. The theory, placed on a generic orientable two-dimensional manifold $\Sigma$, is governed by the action
\begin{align}\label{EQ:JT_action}
  I_{\mathrm{JT}} = -S_0\,I_{\mathrm{EH}} - \frac{1}{2} \int_{\Sigma} \mathrm{d}x^2 \sqrt{g}\,\phi(R + 2) - \int_{\partial\Sigma} \mathrm{d}x\,\sqrt{h}\,\phi(\kappa-1) \;,
\end{align}
where $h$ is the induced metric on the boundary $\partial\Sigma$ and $\kappa$ the extrinsic curvature.
The two-dimensional Einstein--Hilbert action
\begin{align}
  I_{\mathrm{EH}} = \frac{1}{4\pi} \int_{\Sigma} \mathrm{d}x^2 \sqrt{g}\,R + \frac{1}{2\pi}\int_{\partial\Sigma} \mathrm{d}x\,\sqrt{h}\,\kappa
\end{align}
is a purely topological term that computes the Euler characteristic $\chi(\Sigma)$.

In the present work, we consider the Euclidean theory with Dirichlet boundary conditions imposed on the fields: the $n$ connected components of the boundary $\partial\Sigma$ have assigned lengths $\beta_1/\epsilon, \ldots, \beta_n/\epsilon$, while the dilaton is taken to have constant value $\phi_b = \phi_r/\epsilon$ along each boundary component. Such a theory has been extensively studied in the double scaling limit $\epsilon\rightarrow0$, where $\epsilon$ plays the role of a ``holographic renormalization'' parameter. The dilaton field acts as a Lagrange multiplier enforcing the constraint $R=-2$ and thus fixing the bulk geometry. Once integrated out, the nontopological part of the action reduces to a boundary term.

\begin{figure}[htb]
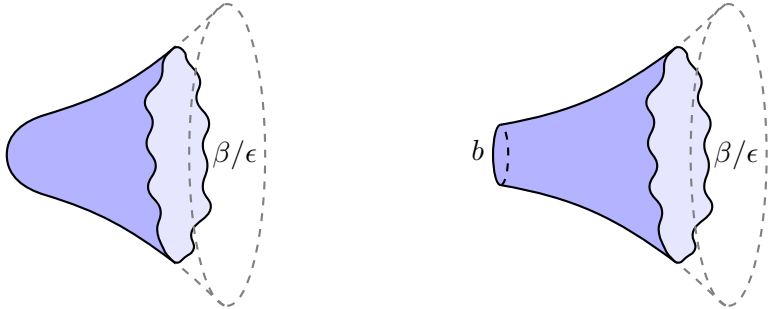

  \centering
  \disk \qquad\qquad\qquad \trumpet
  \caption{\label{FIG:disk_trumpet} The disk (left) and trumpet (right) topologies. The dashed lines represent the full AdS$_2$ geometry, while the actual manifolds have wiggly boundary of length $\beta/\epsilon$. The trumpet has an additional boundary, running along a geodesic of length $b$.}
\end{figure}

The only topology supporting classical solutions is the disk (see Figure~\ref{FIG:disk_trumpet}), where the single boundary cuts out some inner region of Euclidean AdS$_2$ with length $\beta/\epsilon$. In the $\epsilon\rightarrow0$ limit, the theory reduces to the boundary dynamics of a single reparametrization mode $\theta(\xi)$ with Schwarzian action
\begin{equation}\label{EQ:Z_Schwarzian}
  I_{\mathrm{Schw}} = \frac{\phi_r}{2} \int_0^\beta \mathrm{d}\xi \left(\frac{\theta''^2(\xi)}{\theta'^2(\xi)} - \theta'^2(\xi) \right) \;.
\end{equation}
The associated quantum theory is one-loop exact, and its partition function can be written as the Boltzmann integral
\begin{equation}\label{EQ:Z_disk_undeformed}
  Z^{\mathrm{disk}}_{\mathrm{Schw}} = \int_{0}^{\infty} \mathrm{d}E \; \frac{\phi_r\sinh(2\pi\sqrt{2\phi_rE})}{2 \pi^2} \; e^{-\beta E} \;.
\end{equation}

A topology that plays a crucial role in constructing results for generic manifolds is the trumpet (see Figure~\ref{FIG:disk_trumpet}), which differs from the disk by the presence of a geodesic boundary of length $b$. The additional boundary can be taken into account by considering a disk with a hyperbolic defect in the bulk \cite{Mertens:2019tcm}.
This choice leads to a Boltzmann integral with a modified density of states,
\begin{equation}\label{EQ:Z_trumpet_undeformed}
  Z^{\mathrm{trumpet}}_{\mathrm{Schw}} = \int_{0}^{\infty} \mathrm{d}E \; \frac{\phi_r\cos(b\sqrt{2\phi_rE})}{\pi\sqrt{2\phi_rE}} \; e^{-\beta E} \;.
\end{equation}

In \cite{Gross:2019ach} a certain integral deformation of the Schwarzian theory was considered, which is the one-dimensional analogue of the $T\bar{T}$ deformation. It introduces a shift of the energy levels of the theory which is exactly solvable in terms of a parameter $t$. The shift is controlled by the following differential equation for the Hamiltonian $H$,
\begin{align}\label{EQ:TT_def_spectrum}
  2\partial_t H = \frac{\phi_rH^2}{1-\phi_rtH} \;,
\end{align}
which governs the flow of the theory under the deformation.\footnote{The $T\bar{T}$ deformation parameter $t$ is defined is such a way to match the expansion parameter $\phi_b^{-2}$ that we will use for the bulk theory. To match the conventions of \cite{Gross:2019ach,Gross:2019uxi,Iliesiu:2020zld}, one should set $t = 4\lambda/\phi_r$.}
The solutions form two branches
\begin{align}\label{EQ:TT_def_spectrum_branches}
  H_{\pm}(t) = \frac{1}{\phi_rt} \, \left(1\mp\sqrt{1-2\phi_rtE}\right) \;,
\end{align}
however, only $H_+(t)$ reproduces the expected undeformed limit for $t\to 0$. The deformed partition function\footnote{The integrals \eqref{EQ:Z_disk_TT_def_naive} and \eqref{EQ:Z_trumpet_TT_def_naive} are actually ill-defined if the integration region spans the entire positive line. In \cite{Iliesiu:2020zld}, it was suggested to Wick-rotate the bare parameters of the theory to make sense of these integrals. In the following, we shall choose a different approach to address this issue.} is defined by introducing the level shift in \eqref{EQ:Z_disk_undeformed} and \eqref{EQ:Z_trumpet_undeformed},
\begin{align}
  \label{EQ:Z_disk_TT_def_naive}
  Z^{\mathrm{disk}}_{T\bar{T}} &= \int_{0}^{\infty} \mathrm{d}E \; \frac{\phi_r\sinh(2\pi\sqrt{2\phi_rE})}{2 \pi^2} \; e^{-(\beta/\phi_rt)(1-\sqrt{1-2\phi_rtE})} \;, \\
  \label{EQ:Z_trumpet_TT_def_naive}
  Z^{\mathrm{trumpet}}_{T\bar{T}} &= \int_{0}^{\infty} \mathrm{d}E \; \frac{\phi_r\cos(b\sqrt{2\phi_rE})}{\pi\sqrt{2\phi_rE}} \; e^{-(\beta/\phi_rt)(1-\sqrt{1-2\phi_rtE})} \;.
\end{align}
It was then argued \cite{Iliesiu:2020zld} that the deformed Schwarzian theory is the holographic dual of JT gravity at finite cutoff, establishing a correspondence between the $T\bar{T}$ deformation parameter $t$ and the bulk gravity cutoff $\epsilon$. For finite values of the boundary length, as the boundary of $\Sigma$ is pushed away from the asymptotic boundary of AdS$_2$ and into the bulk, the boundary theory flows accordingly.

\section{Disk and trumpet}\label{SEC:disk_trumpet_from_resurgence}
This section is devoted to studying both the disk and the trumpet partition functions for the theory at finite cutoff.

At infinite cutoff, the $\epsilon$ parameter is introduced to take the double-scaling limit where both the boundary length and the value of the dilaton on the boundary diverge, while their ratio $u = \beta/\phi_r$ remains constant. In principle, when considering the theory at finite cutoff, $\epsilon$ becomes redundant since the theory should only depend on its bare parameters. However, as mentioned in Section~\ref{SEC:review}, $\epsilon$ plays an important role, as it parametrizes the deviation from the infinite cutoff limit and is the analog of the deformation parameter $t=\epsilon^2/\phi_r^2$ in the $T\bar{T}$-deformed Schwarzian theory. Moreover, the starting point of our analysis is to study the theory from the point of view of its perturbative expansion in $t$. For this reason, throughout most of the paper, we will find it convenient to express the results in terms $u$ and $t$.

With a simple change of variables, we can recast the disk and the trumpet partition functions in \eqref{EQ:Z_disk_TT_def_naive} and \eqref{EQ:Z_trumpet_TT_def_naive} as
\begin{align}
  \label{EQ:Z_disk_integral}
  Z^{\text{disk}}(u,t) &= \frac{1}{2\pi^2} \int_0^\infty \mathrm{d}s \; s \sinh(2\pi s) \; e^{-I(u,t;s)} \;, \\
  \label{EQ:Z_trumpet_integral}
  Z^{\text{trumpet}}(u,b,t) &= \frac{1}{\pi} \int_0^\infty \mathrm{d}s \; \cos(bs) \; e^{-I(u,t;s)} \;, 
\end{align}
where the $t$-deformed action reads
\begin{align}\label{EQ:action}
  I(u,t;s) = \frac{u}{t}\left(1-\sqrt{1-ts^2}\right) \;.
\end{align}
For any $t>0$, the part of the action \eqref{EQ:action} depending on $s$ becomes imaginary in the region $s\in(1/\sqrt{t},+\infty)$ and the integral diverges. Moreover, the expression above is ambiguous as it is not clear a priori which of the two branches of the square root one should take when crossing the branch point at $s=1/\sqrt{t}$. As we will see in the following, these aspects ultimately signal the presence of non-analytic (namely instanton-like) contributions in the parameter $t$, and care should be taken in choosing the correct prescription to take these into account and make sense of the integrals above.

\subsection{Perturbative expansion}\label{SEC:perturbative_expansion}
Despite the possible ambiguities in defining the integrals \eqref{EQ:Z_disk_integral} and \eqref{EQ:Z_trumpet_integral}, they can be used to yield well-defined asymptotic series in $t$ for the partition functions. This is achieved by first expanding the exponential term as
\begin{align}\label{EQ:A_n_coefficients_decomposition}
  e^{-I(u,t;s)} = e^{-us^2/2} \left(1 + \sum_{n=1}^\infty A_n(s,u) \, t^n\right) \;.
\end{align}
The coefficients $A_n$ can be expressed in terms of generalized Laguerre polynomials,
\begin{align}\label{EQ:A_n_coefficients}
  A_n(s,u) = -\frac{u \, s^{2n+2}}{2^{2n+1}n} \, L_{n-1}^{n+1}\bigg(\frac{us^2}{2}\bigg) \;.
\end{align}
In Appendix \ref{APP:power_expansion}, we show in detail how the expression above is derived.
By integrating each term in the series, we rewrite the partition functions as
\begin{align}
  Z^{\text{disk}}(u,t) &= \sum_{n=0}^\infty Z_{n}^{\text{disk}}(u) \, t^n \;, \label{EQ:t_expansion_Z_disk}\\
  Z^{\text{trumpet}}(u,b,t) &= \sum_{n=0}^\infty Z_{n}^{\text{trumpet}}(u,b) \, t^n \label{EQ:t_expansion_Z_trumpet} \;. 
\end{align}
The $t^0$ terms come from taking the integral against the undeformed action factorized in \eqref{EQ:A_n_coefficients_decomposition} and, as such, correctly reproduce the known results computed in the $t\rightarrow 0$ limit \cite{Saad:2019lba},
\begin{align}
  \label{EQ:zero_t_limit_Z_disk}
  Z_{0}^{\text{disk}}(u) &= \frac{u^{-3/2}}{\sqrt{2\pi}} \; e^{2\pi^2/u} \;, \\
  \label{EQ:zero_t_limit_Z_trumpet}
  Z_{0}^{\text{trumpet}}(u,b) &= \frac{u^{-1/2}}{\sqrt{2\pi}} \; e^{-b^2/2u} \;.
\end{align}
For $n>1$ we have the following expressions:
\begin{align}
  Z_{n}^{\text{disk}}(u) &= \frac{1}{2\pi^2} \int_0^\infty \mathrm{d}s \; s \sinh(2\pi s) \; e^{-us^2/2} \, A_n(s,u) \;, \\
  Z_{n}^{\text{trumpet}}(u,b) &= \frac{1}{\pi} \int_0^\infty \mathrm{d}s \; \cos(bs) \; e^{-us^2/2} \, A_n(s,u) \;.
\end{align}
For any $n\in\mathbb{N}$, the above integrals are real and convergent. By expanding in $t$, we have apparently cured the ambiguity arising from the integration over $s$. We will see in a moment where the subtlety is now hiding. We first need to perform the integration. In order to deal with both integrals at once, we compute
\begin{align}
  a_j
  &= \int_0^\infty \mathrm{d}s \; s^{2j} \; e^{-us^2/2} \, A_n(s,u) \cr
  &= -\frac{1}{n!} \, 2^{j-n-1/2} u^{-j-n-1/2} \bigg(\frac{1}{2}-j\bigg)_{\!n-1} \Gamma\bigg(j+n+\frac{3}{2}\bigg) \;,
\end{align}
were $(x)_n$ denotes the Pochhammer symbol.
Then, we simply use the Taylor expansion of $s \sinh(2\pi s)$ and $\cos(bs)$ to obtain
\begin{align}\label{EQ:Z_n_disk}
  Z_{n}^{\text{disk}}(u)
  &= \frac{1}{2\pi^2} \sum_{j=0}^\infty \frac{(2\pi)^{2j+1} \, a_{j+1}}{(2j+1)!} \cr
  &= Z_{0}^{\text{disk}}(u) \; \frac{(2n)!}{n!} \, (-2u)^{-n} \, L_{2 n}^{3/2-n}\!\left(-\frac{2\pi^2}{u}\right)\cr
  &= \frac{(2u)^{-n}}{n!\sqrt{2\pi^3u^3}} \; \Gamma\bigg(n-\frac{3}{2}\bigg) \, \Gamma\bigg(n+\frac{5}{2}\bigg) \, {}_1F_1\bigg(n+\frac{5}{2};\frac{5}{2}-n;\frac{2\pi^2}{u}\bigg) \;,
\end{align}
and
\begin{align}\label{EQ:Z_n_trumpet}
  Z_{n}^{\text{trumpet}}(u,b)
  &= \frac{1}{\pi} \sum_{j=0}^\infty \frac{(-b^2)^{j} \, a_{j}}{(2j)!} \cr
  &= Z_{0}^{\text{trumpet}}(u) \; \frac{(2n)!}{n!} \, (-2u)^{-n} \, L_{2 n}^{1/2-n}\!\left(\frac{b^2}{2u}\right)\cr
  &= - \frac{(2u)^{-n}}{n!\sqrt{2\pi^3u}} \; \Gamma\bigg(n-\frac{1}{2}\bigg) \, \Gamma\bigg(n+\frac{3}{2}\bigg) \, {}_1F_1\bigg(n+\frac{3}{2};\frac{3}{2}-n;-\frac{b^2}{2 u}\bigg) \;.
\end{align}
Conveniently, the two expressions above capture also the $n=0$ cases in \eqref{EQ:zero_t_limit_Z_disk} and \eqref{EQ:zero_t_limit_Z_trumpet}.
In both cases, the perturbative coefficients are the undeformed partition functions times a polynomial in $1/u$ of degree $3 n$.
Finally, we must remark that these two series expansions can also be directly obtained by solving the flow equation perturbatively in $t$ (see Section~\ref{SEC:flow_equation} for some details) without making any reference to the integral representations \eqref{EQ:Z_disk_integral} and \eqref{EQ:Z_trumpet_integral}.

\subsection{Resurgence}
The coefficients in \eqref{EQ:Z_n_disk} and \eqref{EQ:Z_n_trumpet} grow asymptotically as $n!$. This means that the perturbative expansions in \eqref{EQ:t_expansion_Z_disk} and \eqref{EQ:t_expansion_Z_trumpet} should be understood as formal power series, since both have vanishing radius of convergence. It is possible to associate a finite result to these series by performing a \emph{Borel resummation}.

The Borel sum of a series
\begin{align}\label{EQ:generic_series}
  \Phi(z) = \sum_{n} \omega_n \, z^n 
\end{align}
is defined as follows. First, one should take the \emph{Borel transform} of $\Phi$,
\begin{align}
  \mathcal{B}[\Phi](\zeta) = \sum_{n} \omega_n \, \frac{\zeta^n}{n!} \;.
\end{align}
If the coefficients $\omega_n$ grow as $n!$, $\mathcal{B}[\Phi]$ has finite radius of convergence, thus defining a germ of an analytic function at $\zeta=0$. Then, the directional Borel resummation of $\Phi$ along a chosen direction $\theta$ on the complex $\zeta$-plane is defined as
\begin{align}\label{EQ:Borel_sum}
  \mathcal{S}_{\theta} \Phi(z) = \frac{1}{z} \int_0^{e^{\mathrm{i}\theta}\infty} \mathrm{d}\zeta \; e^{-\zeta/z} \, \mathcal{B}[\Phi](\zeta) \;,
\end{align}
where the integral, taken along the ray with $\arg\zeta = \theta$, is also known as a \emph{directional Laplace transform}. The directional resummation $\mathcal{S}_{\theta} \Phi(z)$ defines an analytic function in the wedge $\operatorname{Re}(e^{-\mathrm{i}\theta}z)>0$ that, upon expansion in $z$, reproduces \eqref{EQ:generic_series}.

In our case, rather than directly computing the Borel transform of \eqref{EQ:t_expansion_Z_disk} and \eqref{EQ:t_expansion_Z_trumpet}, it is convenient to use the power series representation of the Kummer confluent hypergeometric function to first recast \eqref{EQ:Z_n_disk} and \eqref{EQ:Z_n_trumpet} as
\begin{align}
  Z_{n}^{\text{disk}}(u)
  &= \sum_{m=0}^\infty (-2 u)^{-n} \,\frac{u^{-3/2}}{\sqrt{2\pi}} \mathmakebox[\widthof{$\displaystyle\bigg({-}\frac{b^2}{2u}\bigg)^{\!m}$}][c]{\bigg(\frac{2\pi^2}{u}\bigg)^{\!m}} \frac{\Gamma\big(m+n+\frac{5}{2}\big)}{n!m!\, \Gamma\big(m-n+\frac{5}{2}\big)} \;, \\
  Z_{n}^{\text{trumpet}}(u,b)
  &= \sum_{m=0}^\infty (-2 u)^{-n} \,\frac{u^{-1/2}}{\sqrt{2\pi}} \bigg({-}\frac{b^2}{2u}\bigg)^{\!m} \frac{\Gamma\big(m+n+\frac{3}{2}\big)}{n!m!\, \Gamma\big(m-n+\frac{3}{2}\big)} \;,
\end{align}
and then perform the Borel transform on each term in the sum over $m$\footnote{In fact, for fixed $m$, the modulus of the coefficient of the series in $n$ behaves as $\frac{(n-1)!}{\pi m!}$ when $n$ approaches infinity.}. When summing over $n$, each series has finite radius of convergence,
\begin{align}
  \mathcal{B}[Z^{\text{disk}}](u,\zeta) \label{EQ:Borel_transform_disk}
  &= \frac{u^{-3/2}}{\sqrt{2\pi}} \sum_{m=0}^\infty \frac{1}{m!} \mathmakebox[\widthof{$\displaystyle\bigg({-}\frac{b^2}{2u}\bigg)^{\!m}$}][c]{\bigg(\frac{2\pi^2}{u}\bigg)^{\!m}} {}_2F_1\bigg({-}m-\frac{3}{2},m+\frac{5}{2};1;\frac{\zeta }{2 u}\bigg) \;, \\
  \mathcal{B}[Z^{\text{trumpet}}](u,b,\zeta) \label{EQ:Borel_transform_trumpet}
  &= \frac{u^{-1/2}}{\sqrt{2\pi}} \sum_{m=0}^\infty \frac{1}{m!} \bigg({-}\frac{b^2}{2u}\bigg)^{\!m} {}_2F_1\bigg({-}m-\frac{1}{2},m+\frac{3}{2};1;\frac{\zeta }{2 u}\bigg) \;.
\end{align}
Now, in order to complete the Borel summation and obtain an analytic expression for the formal series in \eqref{EQ:t_expansion_Z_disk} and \eqref{EQ:t_expansion_Z_trumpet} one should proceed as in \eqref{EQ:Borel_sum}. However, the hypergeometric functions appearing in the Borel transforms \eqref{EQ:Borel_transform_disk} and \eqref{EQ:Borel_transform_trumpet} have branch cuts located on the positive real axis in the range $\zeta\in(2u, +\infty)$. The branch cut identifies a \emph{Stokes line} at $\arg\zeta = 0$, i.e.\ a singular direction in the $\zeta$ plane. Namely, when taking a directional Laplace transform at $\theta = 0$, one runs into an ambiguity since the results obtained by approaching the Stokes line from above and below differ.

In the theory of resurgence, Stokes lines are associated with nonperturbative contributions, encoded by the discontinuity $(\mathcal{S}_{\theta_{\star}^+} - \mathcal{S}_{\theta_{\star}^-}) \, \Phi(z)$ in the directional Borel resummation as the ray of angle $\theta$ crosses the Stokes line at $\theta_{\star}$. The directional Borel resummations approaching the Stokes lines from both sides are usually referred to as \emph{lateral Borel resummations}. Because of the nonperturbative nature of the discontinuity, both $\mathcal{S}_{\theta_{\star}^+}\Phi(z)$ and $\mathcal{S}_{\theta_{\star}^-}\Phi(z)$ share the same expansion in $z$, but crucially differ by instantonic contributions. In general, the correct nonperturbative completion of $\Phi(z)$ is obtained by choosing some combination of the two. If $\Phi$ is real and the Stokes line lies at $\theta_\star = 0$, under some general assumptions the correct real completion of $\Phi(z)$ is given by the \emph{median resummation}
\begin{align}
  \mathcal{S}_{\mathrm{med}} \Phi(z) = \frac{1}{2}\,(\mathcal{S}_{0^+} + \mathcal{S}_{0^-}) \, \Phi(z) \;.
\end{align}

\begin{figure}[htb]
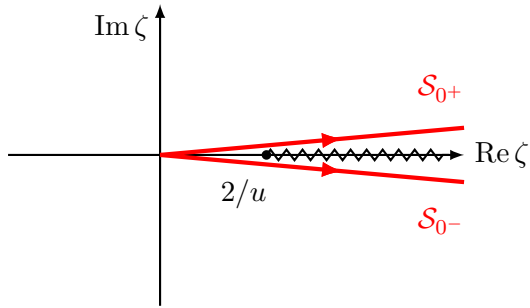

  \centering
  \lateralborel
  \caption{\label{FIG:lateral_borel} The lateral Borel resummations of $Z^{\text{disk}}$ and $Z^{\text{trumpet}}$ approaching the Stokes line at $\theta = 0$ from above and below.}
\end{figure}

Let us apply this to the case at hand. In Appendix \ref{APP:laplace_transform_hypergeometric}, we provide details on how to compute Laplace transforms of Gauss hypergeometric functions above and below the cut. These, in turn, give us the lateral Borel resummations of the disk and trumpet partition functions starting from the expressions for their Borel transforms in \eqref{EQ:Borel_transform_disk} and \eqref{EQ:Borel_transform_trumpet}, 
\begin{align}
  \mathcal{S}_{0^\pm}Z^{\mathrm{disk}}(u,t)
  &= \frac{e^{-\frac{u}{t}}}{\pi u\sqrt{t}} \sum_{m=0}^\infty \frac{1}{m!} \mathmakebox[\widthof{$\displaystyle\bigg({-}\frac{b^2}{2u}\bigg)^{\!m}$}][c]{\bigg(\frac{2\pi^2}{u}\bigg)^{\!m}} \left[\pi\,I_{m+2}\bigg(\frac{u}{t}\bigg) \pm (-1)^{m}\mathrm{i} \, K_{m+2}\bigg(\frac{u}{t}\bigg) \right] \;, \cr
  \mathcal{S}_{0^\pm}Z^{\mathrm{trumpet}}(u,b,t)
  &= \mathmakebox[\widthof{$\displaystyle\frac{e^{-\frac{u}{t}}}{\pi u\sqrt{t}}$}][c]{\frac{e^{-\frac{u}{t}}}{\pi\sqrt{t}}} \sum_{m=0}^\infty \frac{1}{m!} \bigg({-}\frac{b^2}{2u}\bigg)^{\!m} \left[\pi\,I_{m+1}\bigg(\frac{u}{t}\bigg) \pm (-1)^{m}\mathrm{i} \, K_{m+1}\bigg(\frac{u}{t}\bigg) \right] \;,
\end{align}
as depicted in Figure~\ref{FIG:lateral_borel}.
We see that the median resummation mentioned above, indeed, cancels the imaginary terms in the two lateral Borel resummations and gives the real disk and trumpet partition functions
\begin{align}
  \label{EQ:Z_disk}
  Z^{\text{disk}}(u,t) &= \frac{e^{-\frac{u}{t}}}{u\sqrt{t}} \sum_{m=0}^\infty \frac{1}{m!} \mathmakebox[\widthof{$\displaystyle\bigg({-}\frac{b^2}{2u}\bigg)^{\!m}$}][c]{\bigg(\frac{2\pi^2}{u}\bigg)^{\!m}} I_{m+2}\bigg(\frac{u}{t}\bigg) \;, \\
  \label{EQ:Z_trumpet}
  Z^{\text{trumpet}}(u,b,t) &= \mathmakebox[\widthof{$\displaystyle\frac{e^{-\frac{u}{t}}}{u\sqrt{t}}$}][c]{\frac{e^{-\frac{u}{t}}}{\sqrt{t}}} \sum_{m=0}^\infty \frac{1}{m!} \bigg({-}\frac{b^2}{2u}\bigg)^{\!m} I_{m+1}\bigg(\frac{u}{t}\bigg) \;.
\end{align}
Both sums can be performed through the Bessel multiplication theorem
\begin{align}\label{EQ:BesselI_multiplication_theorem}
  \sum_{k=0}^\infty \frac{1}{k!} \left(\frac{z(\lambda^2-1)}{2}\right)^k I_{n+k}(z) &= \lambda^{-n} \, I_n(\lambda z) \;,
\end{align}
to obtain
\begin{align}
  \label{EQ:Z_disk_final}
  Z^{\text{disk}}(u,t)
  &= \frac{u}{\sqrt{t}} \frac{e^{-u/t}}{u^2+4\pi^2t} \, I_2\bigg(\frac{1}{t}\sqrt{u^2+4\pi^2t}\bigg) \;, \\
  \label{EQ:Z_trumpet_final}
  Z^{\text{trumpet}}(u,b,t)
  &= \frac{u}{\sqrt{t}} \frac{e^{-u/t}}{\sqrt{u^2-b^2t}} \, I_1\bigg(\frac{1}{t}\sqrt{u^2-b^2t}\bigg) \;.
\end{align}
Through resurgence, we have been able to unambiguously fix the nonperturbative completions of both the disk and the trumpet partition functions with just their perturbative expansions at $t=0$ as inputs. These corrections naturally carry the information of the nonperturbative branch $H_-$ of the $T\bar{T}$ deformation in \eqref{EQ:TT_def_spectrum_branches}.

\section{The spectrum}\label{SEC:spectrum}
The results for the disk and the trumpet partition functions obtained in \eqref{EQ:Z_disk_final} and \eqref{EQ:Z_trumpet_final} through resurgence can be reproduced by changing the prescription for the integration contour in \eqref{EQ:Z_disk_integral} and \eqref{EQ:Z_trumpet_integral}. At the level of the boundary theory, this prescription induces a cutoff on the spectrum for any finite value of $t$ and gives rise to instantonic contributions associated with a region in the spectrum where the density of states becomes negative. The present section is dedicated to discussing these aspects.

\subsection{Integration contour}
\begin{figure}[htb]
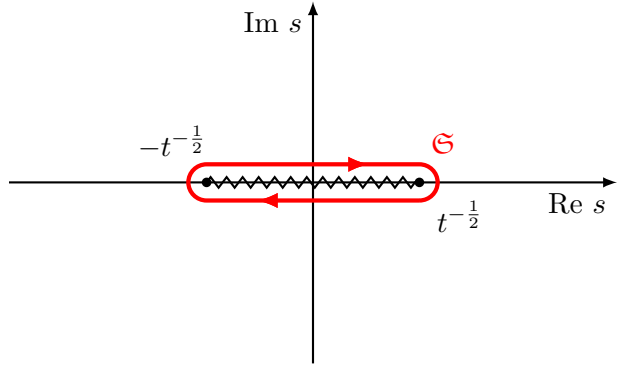

  \centering
  \dogbone
  \caption{\label{FIG:dogbone} The contour $\mathfrak{S}$ surrounding the branch cut of the integrands in \eqref{EQ:Z_disk_integral} and \eqref{EQ:Z_trumpet_integral}.}
\end{figure}

The action \eqref{EQ:action} has two branch points, located at $s = -1/\sqrt{t}$ and $s = +1/\sqrt{t}$. We can extend the definition of $I(t,u;s)$ to the complex $s$-plane by placing a branch cut in the interval $s\in(-1/\sqrt{t},+1/\sqrt{t})$. Then, we can replace the original contour, running along the real axis, with an integration contour $\mathfrak{S}$ surrounding the branch as depicted in Figure~\ref{FIG:dogbone}.

Let us consider a generic integral
\begin{align}
  W = \int_\mathfrak{S} \mathrm{d}s \; f(s) \, e^{-I(t,u;s)} \;,
\end{align}
where $f$ is some entire function. The integral is easily computed in terms of the discontinuity of the action across the branch cut,
\begin{align}
  W 
  &= 2e^{-u/t}\int_{-1/\sqrt{t}}^{+1/\sqrt{t}} \mathrm{d}s \; f(s) \, \sinh\!\left(\frac{u}{t}\sqrt{1-ts^2}\right) \cr
  &= \frac{2e^{-u/t}}{\sqrt{t}}\int_{0}^{\pi} \mathrm{d}\theta \; \sin\theta \, f\bigg(\frac{\cos\theta}{\sqrt{t}}\bigg) \, \sinh\!\bigg(\frac{u}{t}\sin\theta\bigg) \;,
\end{align}
where in the last step we introduced the change of variable $\cos\theta = \sqrt{t}s$.
By replacing the hyperbolic sine with its Taylor expansion, we find
\begin{align}
  W
  &= \sum_{j=0}^\infty \bigg(\frac{u}{t}\bigg)^{\!2j+1} \frac{2e^{-u/t}}{(2j+1)!\sqrt{t}}\int_{0}^{\pi} \mathrm{d}\theta \; (\sin\theta)^{2j+2} \, f\bigg(\frac{\cos\theta}{\sqrt{t}}\bigg) \;.
\end{align}

\paragraph{Disk.}
The disk partition function is obtained by setting $f(s) = s \sinh(2\pi s)/(4\pi^2)$. Actually, because the original integral is even in $s$ and the integration range is symmetric about the origin, one can equivalently use $f(s) = s \exp(2\pi s)/(4\pi^2)$, which gives
\begin{align}
  Z^{\text{disk}}(u,t) &= \sum_{j=0}^\infty \bigg(\frac{u}{t}\bigg)^{\!2j+1} \frac{e^{-u/t}}{2\pi^2(2j+1)!t}\int_{0}^{\pi} \mathrm{d}\theta \; (\sin\theta)^{2j+2} \, \cos\theta \, \exp\!\bigg(\frac{2\pi\cos\theta}{\sqrt{t}}\bigg) \;.
\end{align}
The term
\begin{align}
  (\sin\theta)^{2j+2} \, \cos\theta = \frac{1}{2j+3} \, \frac{\mathrm{d}}{\mathrm{d}\theta}(\sin\theta)^{2j+3}
\end{align}
can be used to integrate by parts and obtain
\begin{align}
  Z^{\text{disk}}(u,t) &= \sum_{j=0}^\infty \bigg(\frac{u}{t}\bigg)^{\!2j+1} \frac{e^{-u/t}}{\pi(2j+1)!(2j+3)t^{3/2}}\int_{0}^{\pi} \mathrm{d}\theta \; (\sin\theta)^{2j+4} \, \exp\!\bigg(\frac{2\pi\cos\theta}{\sqrt{t}}\bigg) \;.
\end{align}
We perform the integration by using the integral representation of the modified Bessel function, that, for $j\in\mathbb{Z}$, reads
\begin{align}
  I_{j}(z) = \frac{z^{j}2^{j}j!}{(2j)!\pi}\int_{0}^{\pi} \mathrm{d}\theta \; (\sin\theta)^{2j} \, \exp(z\cos\theta) \;.
\end{align}
This gives
\begin{align}
  Z^{\text{disk}}(u,t) &= \frac{u e^{-\frac{u}{t}}}{4\pi^2t^{3/2}} \sum_{j=0}^{\infty} \frac{1}{j!} \left(\frac{u^2}{4 \pi t^{3/2}}\right)^j I_{j+2}\left(\frac{2\pi}{\sqrt{t}}\right) \;,
\end{align}
which, upon summation with \eqref{EQ:BesselI_multiplication_theorem}, reproduces the result computed in \eqref{EQ:Z_disk_final}.

\paragraph{Trumpet.}
Likewise, the trumpet partition function is recovered from $W$ by setting $f(s) = \cos(bs)/(2\pi)$,
\begin{align}
  Z^{\text{trumpet}}(u,b,t)
  &= \sum_{j=0}^\infty \bigg(\frac{u}{t}\bigg)^{\!2j+1} \frac{e^{-u/t}}{\pi(2j+1)!\sqrt{t}}\int_{0}^{\pi} \mathrm{d}\theta \; (\sin\theta)^{2j+2} \, \cos\bigg(\frac{b\cos\theta}{\sqrt{t}}\bigg) \;.
\end{align}
We use the integral representation
\begin{align}
  J_{j}(z) &= \frac{z^{j}2^{j}j!}{\left(2j\right)!\pi} \int_{0}^{\pi} \mathrm{d}\theta \; (\sin\theta)^{2j} \, \cos(z\cos\theta) \;,
\end{align}
to find
\begin{align}\label{EQ:Z_trumpet_sum_representation}
  Z^{\text{trumpet}}(u,b,t)
  &= \frac{u e^{-\frac{u}{t}}}{bt} \sum_{j=0}^{\infty} \frac{1}{j!} \left(\frac{u^2}{2bt^{3/2}}\right)^j J_{j+1}\left(\frac{b}{\sqrt{t}}\right) \;.
\end{align}
Using the multiplication theorem
\begin{align}
  \sum_{k=0}^\infty \frac{1}{k!} \left(\frac{z(1-\lambda^2)}{2}\right)^k J_{n+k}(z) &= \lambda^{-n} \, J_n(\lambda z) \;,
\end{align}
we finally get
\begin{align}
  Z^{\text{trumpet}}(u,b,t) &= \frac{u}{\sqrt{t}}\frac{e^{-u/t}}{\sqrt{b^2t-u^2}} \, J_1\!\left(\frac{\sqrt{b^2t-u^2}}{t}\right) \;,
\end{align}
which agrees with \eqref{EQ:Z_trumpet_final}, since $I_n(x) = \mathrm{i}^{-n}\,J_n(\mathrm{i}x)$.

\subsection{The density of states}\label{SEC:density_of_states}
The prescription on the integration contour $\mathfrak{S}$ translates into a prescription for the integration over the spectrum of the boundary theory.
Changing the integration variable to $\mathcal{E}=s^2$ brings the disk partition function into the form
\begin{align}
  Z^{\text{disk}}(u,t)
  &= \int_{0}^{1/t} \mathrm{d}\mathcal{E} \; \frac{\sinh\!\big(2\pi\sqrt{\mathcal{E}}\big)}{4\pi^2} \; \Big(e^{-u(1-\sqrt{1-t\mathcal{E}})/t} - e^{-u(1+\sqrt{1-t\mathcal{E}})/t} \Big) \;.
\end{align}
The above differs from the na\"ive integral \eqref{EQ:Z_disk_TT_def_naive} associated with the $T\bar{T}$-deformed Schwarzian theory in two ways: the integration range is now capped at $\mathcal{E}=1/t$, and there is an additional term of instantonic origin. The spectrum is always real within the integration range, and in the $t\rightarrow 0$ limit, one recovers the undeformed Schwarzian partition function.
	
The deformed density of states $\rho(E;t)$ can be obtained as follows. We split the exponential terms into two separate integrals and apply on both an appropriate change of variables to obtain integrals of the type
\begin{align}
  \int \mathrm{d}E \; \rho(E;t) \; e^{-\phi_ruE} \;,
\end{align}
where the density of states is weighted by a Boltzmann factor with inverse temperature $\beta = \phi_ru$. This amounts to set
\begin{align}
  E &= -\frac{1}{\phi_rt}\left(\pm\sqrt{1-t\mathcal{E}}-1\right) \;, & \mathcal{E} &= \frac{1}{t}\,\Big(1-(1-\phi_rtE)^2\Big) \;.
\end{align}
The the two integrals combine nicely as
\begin{align}
  Z^{\text{disk}}(u,t) &= \int_{0}^{1/(\phi_rt)} \mathrm{d}E \; \rho(E;t) \; e^{-\phi_ruE} - \int_{2/(\phi_rt)}^{1/(\phi_rt)} \mathrm{d}E \; \rho(E;t) \; e^{-\phi_ruE} \cr
  &= \int_{0}^{2/(\phi_rt)} \mathrm{d}E \; \rho(E;t) \; e^{-\phi_ruE} \;,
\end{align}
where the $t$-deformed density of states is given by
\begin{align}
  \rho(E;t)
  &= \frac{1}{4\pi^2} \, \sinh\!\left(2\pi\sqrt{\mathcal{E}(E)}\right) \, \frac{\mathrm{d}\mathcal{E}(E)}{\mathrm{d}E} \cr
  &= \frac{\phi_r(1-t\phi_rE)}{2\pi^2}\,\sinh\!\left(2\pi\sqrt{\phi_rE(2-t\phi_rE)}\right) \;.
\end{align}
Here and in the following, whenever we Laplace-transform from $u$ to $E$, we adopt the widely-used convention of setting $\phi_r = 1/2$. With this choice, the density of states reads
\begin{align}\label{EQ:density_of_states}
  \rho(E;t)
  &= \frac{1-tE/2}{4\pi^2}\,\sinh\!\left(2\pi\sqrt{E(1-tE/4)}\right) \;.
\end{align}
To interpret the integral as a conventional Laplace transform we simply extend the integration range to the entire real positive $E$ axis and define $\rho(E;t)$ to have support on the interval $E\in(0, 4/t)$. At $t=0$, the above reproduces the familiar Schwarzian density $\rho\propto\sinh(2\pi\sqrt{E})$ growing exponentially in $\sqrt{E}$. At finite $t$, the result is qualitatively rather different. In the ``perturbative range'' $0<E<2/t$ the density is positive, but after an initial growth it decreases and reaches a zero at $E=2/t$. In the ``nonperturbative range'' $2/t<E<4/t$ the density becomes negative; the two ranges are related by the symmetry property $\rho(4/t-E,t) = -\rho(E;t)$ and thus the integral of $\rho(E;t)$ over the entire spectrum vanishes.

\section{Other topologies}\label{SEC:other_topologies}
The disk partition function $Z^{\text{disk}}(u,t)$ computed in the previous sections is the partition function associated with a manifold of genus zero whose boundary has a single connected component of (rescaled) length $u$,
\begin{align}
  Z_{0,1}(u;t) &= Z^{\text{disk}}(u,t) \;.
\end{align}
In general, one can compute partition functions on orientable manifolds with arbitrary topology. These are classified by the number $n$ of connected components of the boundary, and by the genus $g$. The resulting partition function, $Z_{g,n}$ will depend on the lengths $u_1, \ldots, u_n$ of the connected boundaries.

In a theory of quantum gravity, the path integral receives contributions from different spacetime topologies. Such a property sometimes goes under the name of ``third quantization''. This means that, for any given choice of $n$, the full partition function should really be a sum over the $Z_{g,n}$ obtained for any value of the genus $g$. Each term is weighted by the topological (Einstein--Hilbert) action term that gives a factor of $(e^{S_0})^\chi$, where $\chi=2-2g-n$ is the Euler characteristic. At fixed $n$ the full partition function reads
\begin{align}
  \mathbf{Z}_n(u_1,\ldots,u_n;t) = e^{(2-n)S_0} \sum_{g=0}^\infty \, e^{-2gS_0} \; Z_{g,n}(u_1,\ldots,u_n;t) \;.
\end{align}

In \cite{Saad:2019lba}, it was shown that the partition function $Z_{g,n}$ for a generic choice of $n$ and $g$ can be obtained in terms of a certain topological decomposition. Each boundary component of length $u_i$ is associated with a trumpet $Z^{\text{trumpet}}(u_i,b_i,t)$ that is glued to a bordered Riemann surface of genus $g$ through a common geodesic boundary of length $b_i$. In Figure~\ref{FIG:genusone}, we show the simple case of $Z_{1,1}$. The gluing is performed by taking an integral over the length $b_i$ of each geodesic boundary,
\begin{align}\label{EQ:higher_genus}
  Z_{g,n}(u_1,\ldots,u_n;t) &= \int_0^\infty \mathrm{d}b_1 \; b_1 \ldots \int_0^\infty \mathrm{d}b_n \; b_n \, V_{g,n}(b_1,\ldots,b_n) \cr
  &\qquad\qquad \times Z^{\text{trumpet}}(u_1,b_1,t) \ldots Z^{\text{trumpet}}(u_n,b_n,t) \;.
\end{align}
The formula is written in terms of $V_{g,n}(b_1,\ldots,b_n)$, the Weil--Petersson volume of the moduli space of hyperbolic Riemann surfaces of genus $g$ with $n$ geodesic boundaries of lengths $b_1,\ldots, b_n$. The Weil--Petersson volume can be represented as an integral over the Deligne--Mumford compactification $\overline{\mathcal{M}}_{g,n}$ of the moduli space of Riemann surfaces of genus $g$ and $n$ marked points $p_i$,
\begin{align}\label{EQ:Weil-Petersson_integral}
  V_{g,n}(b_1,\ldots,b_n) &= \int_{\overline{\mathcal{M}}_{g,n}} \exp\bigg(\omega + \frac{1}{2}\sum_{i=1}^n\psi_i^{\vphantom{2}}\,b_i^2\bigg) \;,
\end{align}
where $\omega$ is the Weil--Petersson symplectic form and $\psi_i$ the first Chern class of the line bundle whose fiber is the cotangent space at $p_i$. We refer the reader to \cite{Dijkgraaf:2018vnm} for an introduction on the subject.

\begin{figure}[htb]
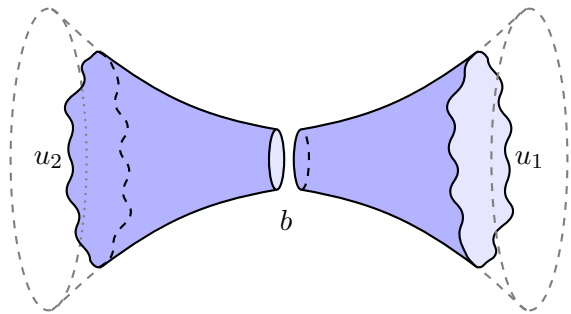

  \centering
  \cylinder
  \caption{\label{FIG:cylinder} The topological decomposition of the cylinder in terms of two trumpets glued along their geodesic boundary.}
\end{figure}

Besides the disk, the only other topology that represents an exception to the formula above is given by the cylinder
\begin{align}\label{EQ:cylinder_def}
  Z_{0,2}(u_1,u_2;t) &= \int_0^\infty \mathrm{d}b \; b \, Z^{\text{trumpet}}(u_1,b,t) \, Z^{\text{trumpet}}(u_2,b,t) \;,
\end{align}
which is obtained by directly gluing together two trumpets along their geodesic boundary, as shown in Figure~\ref{FIG:cylinder}.

In principle, it is not obvious why the gluing prescription should still be valid at finite $t$. In fact, for any $t>0$, there is a portion of the integration range where the length $b$ of the geodesic boundary exceeds the length $u/\sqrt{t}$ of the Dirichlet boundary of the same trumpet. However, as already noticed in \cite{Iliesiu:2020zld}, the inclusion of nonperturbative terms had the effect of making the trumpet partition function in \eqref{EQ:Z_trumpet_final} regular at $b = u/\sqrt{t}$ and real across the entire integration range, thus making the integral well-defined. Although we do not have an \emph{ab initio} derivation of \eqref{EQ:higher_genus} and \eqref{EQ:cylinder_def} for finite $t$, we take these formulas as prescriptions for the computation of partition functions for any topology. We will show later that the results they generate have remarkable properties.

\subsection{Cylinder}
We start our analysis of by considering the cylinder $Z_{0,2}(u_1,u_2;t)$. The series representation in \eqref{EQ:Z_trumpet_sum_representation} turns out to be particularly useful when dealing with integrals over $b$. We plug that expression in \eqref{EQ:cylinder_def} and use
\begin{align}
  \int_0^\infty \frac{\mathrm{d} b}{b^{j+k+1}} \, J_{j+1}\bigg(\frac{b}{\sqrt{t}}\bigg) \, J_{k+1}\bigg(\frac{b}{\sqrt{t}}\bigg)
  &= \frac{1}{2j!k!(j+k+1)} \bigg(\frac{1}{2\sqrt{t}}\bigg)^{j+k}
\end{align}
to write
\begin{align}
  Z_{0,2}(u_1,u_2;t)
  &= 2u_1u_2 \, e^{-(u_1+u_2)/t} \, \sum_{j=0}^\infty\sum_{k=0}^\infty \frac{u_1^{2j}u_2^{2k}}{(j!)^2(k!)^2} \bigg(\frac{1}{4t^2}\bigg)^{j+k+1} \frac{1}{(j+k+1)} \;.
\end{align}
To split the sums, we can use the trivial identity
\begin{align}
  \bigg(\frac{1}{4t^2}\bigg)^{j+k+1} \frac{1}{(j+k+1)} = \int_0^{1/(2t)^2} \mathrm{d}x \; x^{j+k}
\end{align}
and perform the summations
\begin{align}
  \sum_{j=0}^\infty\frac{(xu^2)^j}{(j!)^2} = I_0(2\sqrt{x}u)
\end{align}
to obtain
\begin{align}\label{EQ:cylinder_last_step}
  Z_{0,2}(u_1,u_2;t)
  &= 2u_1u_2 \, e^{-(u_1+u_2)/t} \, \int_0^{1/(2t)^2} \mathrm{d} x \; I_0(2\sqrt{x}u_1) \, I_0(2\sqrt{x}u_2) \;.
\end{align}
The last integral is easily evaluated and gives
\begin{align}\label{EQ:cylinder}
  Z_{0,2}(u_1,u_2;t)
  &= \frac{u_1u_2 \, e^{-(u_1+u_2)/t}}{t(u_1^2-u_2^2)} \, \left[u_1\,I_0\!\left(\frac{u_2}{t}\right)\,I_1\!\left(\frac{u_1}{t}\right) - u_2\,I_0\!\left(\frac{u_1}{t}\right)\,I_1\!\left(\frac{u_2}{t}\right)\right] \;.
\end{align}
It is not immediate to see how this extends the result at infinite cutoff, but performing an expansion at the first few orders in $t$, one finds
\begin{align}
  Z_{0,2}(u_1,u_2;t)
  &= \frac{1}{2\pi}\frac{\sqrt{u_1}\sqrt{u_2}}{(u_1+u_2)} + \frac{1}{26\pi}\frac{1}{\sqrt{u_1}\sqrt{u_2}}\,t + \frac{9}{256\pi}\frac{u_1+u_2}{(\sqrt{u_1}\sqrt{u_2})^3}\,t^2 + O(t^3) \;,
\end{align}
which matches the $t\rightarrow 0$ limit previously computed in \cite{Saad:2019lba}. However, a power expansion in $t$ necessarily misses nonperturbative contributions that, as we will discuss later, constitute a crucial feature of the theory at finite cutoff.

From the cylinder partition function one can extract the resolvent by taking a Laplace transform over both $u_1$ and $u_2$,\footnote{
Here, we make use of the trivial change of variables
\begin{align}
  -\int_0^\infty \mathrm{d}\beta \; e^{-\beta E} f(\beta) = -\phi_r\int_0^\infty \mathrm{d}u \; e^{-\phi_ruE} f(\phi_ru) \;,
\end{align}
together with the convention, stated at the end of Section~\ref{SEC:density_of_states}, according to which $\phi_r=1/2$.
}
\begin{align}
  R_{0,2}(E_1,E_2;t) = \frac{1}{4}\int_0^\infty\!\mathrm{d}u_1 \int_0^\infty\!\mathrm{d}u_2 \; Z_{0,2}(u_1,u_2;t) \; e^{u_1E_1/2 + u_2E_2/2} \;.
\end{align}
We assume $E_1, E_2 < 0$ and apply the identity
\begin{align}
  \int_0^\infty \mathrm{d}u \; u \, e^{-\alpha u} \, I_0(\beta u) = \frac{\alpha}{(\alpha^2-\beta^2)^{3/2}} \;,
\end{align}
which holds for $\operatorname{Re}\alpha > |\operatorname{Re}\beta\,|$, to \eqref{EQ:cylinder_last_step}. This gives
\begin{align}\label{EQ:resolvent}
  R_{0,2}(E_1,E_2;t)
  &= \frac{1}{2} \int_0^{1/(2t)^2} \mathrm{d} x \; \frac{1/t-E_1/2}{[(1/t-E_1/2)^2-4x]^{3/2}} \, \frac{1/t-E_2/2}{[(1/t-E_2/2)^2-4x]^{3/2}} \cr
  &= \frac{t^2 (1-tE_1/2) (1-tE_2/2) (tE_1^2/4 + tE_2^2/4 - E_1 - E_2)}{4[(1-tE_1/2)^2-(1-tE_2/2)^2]^2 \sqrt{-E_1(1-tE_1/4)}\sqrt{-E_2(1-tE_2/4)}} \cr
  &\qquad -\frac{t^2 [(1-tE_1/2)^2+(1-tE_2/2)^2]}{4[(1-tE_1/2)^2-(1-tE_2/2)^2]^2} \;.
\end{align}
When continuing the resolvent to arbitrary complex values of $E_1$ and $E_2$, the square roots in \eqref{EQ:resolvent} induce branch cuts at $E_1\in(0,4/t)$ and $E_2\in(0,4/t)$. The double-discontinuity of the resolvent across the real $E_1$ and $E_2$ lines, appropriately normalized by a $-1/(4\pi^2)$ factor, gives the two-point correlator of the density $\rho(E;t)$,
\begin{align}
  \rho_{0,2}(E_1,E_2;t)
  &= \frac{t^2 (1-tE_1/2) (1-tE_2/2) (tE_1^2/4 + tE_2^2/4 - E_1 - E_2)}{4\pi^2[(1-tE_1/2)^2-(1-tE_2/2)^2]^2 \sqrt{E_1E_2(1-tE_1/4)(1-tE_2/4)}} \;.
\end{align}
Interestingly, its support coincides with the one computed for the one-point function of $\rho(E;t)$ in \eqref{EQ:density_of_states}, obtained from the disk partition function. In fact, the above expression is valid within the ranges where the branch cuts extend, i.e.\ for $E_1\in(0,4/t)$ or $E_1\in(0,4/t)$. Outside those ranges, $\rho_{0,2}$ vanishes, since the double discontinuity of the resolvent $R_{0,2}$ is zero.

\subsection{The general case}\label{SEC:general_case}
The Weil--Petersson volume $V_{g,n}$ is a polynomial of degree $3g-3+n$ in the squared lengths $b_1^2, \ldots, b_n^2$ of the geodesic boundaries. By linearity, \eqref{EQ:higher_genus} can be computed by splitting the contribution of each monomial as a product of integrals where a single $Z^{\mathrm{trumpet}}(u,b,t)$ is integrated against some even power of $b$.
It is sufficient to use
\begin{align}\label{EQ:Z_building_block}
  \tilde{Z}_m(u,t)
  &= \int_0^\infty \mathrm{d}b \; b \, Z^{\mathrm{trumpet}}(u,b,t) \, b^{2m} \cr
  &= \frac{m!}{\sqrt{t}} \, 2^mu^{m+1} e^{-u/t} \, I_{m}\!\left(\frac{u}{t}\right)
\end{align}
to read off $Z_{g,n}$ from the coefficients in the polynomial $V_{g,n}$. Moreover, we can use the representation of $V_{g,n}$ in \eqref{EQ:Weil-Petersson_integral} as an integral over $\overline{\mathcal{M}}_{g,n}$ together with \eqref{EQ:Z_building_block} to recast \eqref{EQ:higher_genus} as\footnote{The infinite sum can be rewritten in terms of Lommel functions of two variables as
\begin{align}
  \sum_{j=0}^\infty (\psi_i\,u_i)^j \, I_j\!\left(\frac{u_i}{t}\right) = V_0\!\left(\frac{\mathrm{i}}{t\psi_i},\frac{\mathrm{i}u_i}{t}\right) - \mathrm{i}V_1\!\left(\frac{\mathrm{i}}{t\psi_i},\frac{\mathrm{i}u_i}{t}\right) \;.
\end{align}}
\begin{align}
  Z_{g,n}(u_1,\ldots,u_n;t) &= \frac{1}{t^{n/2}} \int_{\overline{\mathcal{M}}_{g,n}} e^{\omega} \prod_{i=1}^n u_i \, e^{-u_i/t} \sum_{j=0}^\infty (\psi_i\,u_i)^j \, I_j\!\left(\frac{u_i}{t}\right) \;.
\end{align}

A similar approach can be used to compute generic resolvents $R_{g,n}(E_1,\ldots,E_n;t)$. Instead of taking an integral transform of the result $Z_{g,n}(u_1,\ldots,u_n;t)$, it is possible to first apply the transformation to a single trumpet. This generates a term, for $E<0$,
\begin{align}
  T(E,b,t)
  &= -\frac{1}{2}\int_0^\infty \mathrm{d}u \; Z^{\mathrm{trumpet}}(u,b,t) \, e^{uE/2} \cr
  &= -\frac{\sqrt{t}}{2\sqrt{\pi}} \sum_{k=1}^\infty \frac{\Gamma(k+1/2)}{(1-tE/2)^{2k}} \bigg(\frac{2\sqrt{t}}{b}\bigg)^{\!k} \, J_{k}\!\left(\frac{b}{\sqrt{t}}\right)
\end{align}
which can be glued to different topologies as in \eqref{EQ:higher_genus} to compute directly the resolvent for any $g$ and $n$,
\begin{align}
  R_{g,n}(E_1,\ldots,E_n;t) &= \int_0^\infty \mathrm{d}b_1 \; b_1 \ldots \int_0^\infty \mathrm{d}b_n \; b_n \, V_{g,n}(b_1,\ldots,b_n) \; T(E_1,b_1,t) \ldots T(E_n,b_n,t) \;.
\end{align}
Again, because of linearity, one simply needs to use
\begin{align}
  \tilde{R}_m(E,t)
  &= \int_0^\infty \mathrm{d}b \; b \; T(E,b,t) \; b^{2m} \cr
  &= -\frac{(2m+1)!\,(1-tE/2)}{2(-E(1-tE/4))^{m+1}\sqrt{-E(1-tE/4)}} \;,
\end{align}
to immediately obtain the result for $R_{g,n}$ for any given $V_{g,n}$. To compute arbitrary correlators of $\rho(E;t)$ one can take the discontinuity
\begin{align}
  \tilde{\rho}_m(E,t)
  &= \frac{\tilde{R}_m(E-\mathrm{i}0,t)-\tilde{R}_m(E+\mathrm{i}0,t)}{2\pi\mathrm{i}} \cr
  &= (-1)^{m+1} \frac{(2m+1)!\,(1-tE/2)}{2\pi(E(1-tE/4))^{m+1}\sqrt{E(1-tE/4)}} \, \theta(E) \, \theta(4-tE) \;.
\end{align}

\begin{figure}[htb]
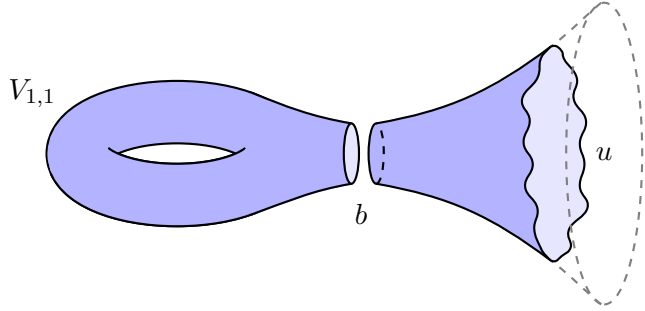

  \centering
  \genusone
  \caption{\label{FIG:genusone} The topological decomposition of a disk at genus one. A trumpet is glued to genus-one Riemann surface along a common geodesic boundary.}
\end{figure}

\paragraph{Examples.} As an example of the above setup, we compute the partition function for the disk at genus one (see Figure~\ref{FIG:genusone}).
The relevant Weil--Petersson volume is given by
\begin{align}
  V_{1,1}(b) &= \frac{1}{48}(4\pi^2+b^2) \;.
\end{align}
By replacing each power of $b$ with the appropriate term in \eqref{EQ:Z_building_block}, we find
\begin{align}
  Z_{1,1}(u;t)
  &= \frac{1}{48}(4\pi^2\tilde{Z}_0(u,t)+\tilde{Z}_2(u,t)) \cr
  &= \frac{u \, e^{-u/t} }{24\sqrt{t}} \left[2 \pi^2 \, I_0\!\left(\frac{u}{t}\right) + u \, I_1\!\left(\frac{u}{t}\right) \right] \;.
\end{align}
When computing results at higher $g$ and $n$, which involve higher powers of $b$, one can recursively apply the identity
\begin{align}
  I_{n+1}(z) = I_{n-1}(z) - \frac{2n}{z} I_{n}(z) \;,
\end{align}
to rewrite the result solely in terms of modified Bessel functions of order zero and one.
For instance, from
\begin{align}
  V_{2,1}(b) &= \frac{1}{2211840}(4\pi^2+b^2)(12\pi^2+b^2)(6960\pi^4+384\pi^2b^2+5b^4) \;, \\
  V_{1,2}(b_1,b_2) &= \frac{1}{192}(4\pi^2+b_1^2+b_2^2)(12\pi^2+b_1^2+b_2^2) \;, \\
  V_{0,3}(b_1,b_2,b_3) &= \vphantom{\frac{1}{1}}1 \;,
\end{align}
we can compute the disk at genus two,
\begin{align}
  Z_{2,1}(u;t)
  &= \frac{u \, e^{-u/t} }{5760\sqrt{t}} (870\pi^8+278\pi^4u^2-232\pi^2tu^2+120t^2u^2+5u^4) \, I_0\!\left(\frac{u}{t}\right) \cr
  &\quad \; + \frac{u \, e^{-u/t} }{2880\sqrt{t}} (338\pi^6-278\pi^4t+232\pi^2t^2-120t^3+29\pi^2u^2-20tu^2) \, I_1\!\left(\frac{u}{t}\right) \;, \cr
\end{align}
the cylinder at genus one,
\begin{align}
  Z_{1,2}(u_1,u_2;t)
  = {}&\frac{u_1 u_2}{24t} \, e^{-(u_1+u_2)/t}
  \left[u_1\!\left(2 \pi ^2- t\right) I_1\!\left(\frac{u_1}{t}\right)I_0\!\left(\frac{u_2}{t}\right) + (u_1 \leftrightarrow u_2) \right. \cr
  &\left. + \left(u_1^2+u_2^2+6 \pi ^4\right) I_0\!\left(\frac{u_1}{t}\right)I_0\!\left(\frac{u_2}{t}\right) + u_1u_2 \, I_1\!\left(\frac{u_1}{t}\right)I_1\!\left(\frac{u_2}{t}\right)\right] \;,
\end{align}
and the topology with three boundaries at genus zero
\begin{align}
  Z_{0,3}(u_1,u_2,u_3;t) &= \frac{u_1 u_2 u_3}{t^{3/2}} \, e^{-(u_1+u_2+u_3)/t} \, I_0\!\left(\frac{u_1}{t}\right) \, I_0\!\left(\frac{u_2}{t}\right) \, I_0\!\left(\frac{u_3}{t}\right) \;.
\end{align}

\section{Flow equation}\label{SEC:flow_equation}
The shift in the spectrum of the $T\bar{T}$-deformed theory as a function of the deformation parameter is controlled by \eqref{EQ:TT_def_spectrum}. The equation imposes a constraint at the level of the thermal partition function, in the form of a partial differential equation in both the deformation parameter $t$ and the (rescaled) temperature $u$. In fact, the differential operator
\begin{align}
  \mathbf{F}(u,t) = u\,\frac{\partial^2}{\partial u^2} + 2t\,\frac{\partial^2}{\partial u\,\partial t} - 2 \left(\frac{t}{u}-1\right) \frac{\partial}{\partial t}
\end{align}
has the property that, for any density of states $\varrho(\mathcal{E})$,
\begin{align}\label{EQ:flow_equation_naive}
  \mathbf{F}(u,t) \int_0^\infty \mathrm{d}\mathcal{E} \; \varrho(\mathcal{E}) \; e^{-u(1-\sqrt{1-t\mathcal{E}})/t} = 0 \;.
\end{align}
This induces a recursion relation
\begin{align}
  2u(n+1) \, Z^{\vphantom{'}}_{n+1}\vphantom{'}(u) = 2n\, Z^{\vphantom{'}}_n(u) - 2nu\, Z_n'(u) - u^2 Z_n''(u)
\end{align}
for both the disk and the trumpet coefficients in the $t$-expansion introduced in Section \ref{SEC:perturbative_expansion}. A simple check on the explicit forms derived in \eqref{EQ:Z_n_disk} and \eqref{EQ:Z_n_trumpet} shows that, indeed, the above holds true.

With the introduction of nonperturbative corrections, however, the integral of the type in \eqref{EQ:flow_equation_naive} should be modified according to the prescriptions discussed in Section \ref{SEC:density_of_states}. Interestingly, the modified integral is still a solution of the flow equation,
\begin{align}
  \mathbf{F}(u,t) \int_{0}^{1/t} \mathrm{d}\mathcal{E} \; \varrho(\mathcal{E}) \; \Big(e^{-u(1-\sqrt{1-t\mathcal{E}})/t} - e^{-u(1+\sqrt{1-t\mathcal{E}})/t} \Big) = 0 \;.
\end{align}
As a consequence, both \eqref{EQ:Z_disk_final} and \eqref{EQ:Z_trumpet_final} are solutions of the flow equation, as one can explicitly check. The nonperturbative contributions to both the disk and the trumpet partition function correct the perturbative series with the addition of a trans-series term of the form 
\begin{align}
  Z_{\mathrm{inst.}}(u,t) = e^{-2u/t} \sum_{n=0}^{\infty} Z_{n}^{*}(u) \, t^n \;.
\end{align}
The presence of the exponential associated with the instantonic saddle has the effect of modifying the action of the flow equation operator by flipping the sign of $u$,
\begin{align}
  \mathbf{F}(u,t) \, Z_{\mathrm{inst.}}(u,t) = - e^{-2u/t} \; \mathbf{F}(-u,t) \sum_{n=0}^{\infty} Z_{n}^{*}(u) \;.
\end{align}
As a consequence, the coefficients $Z_n^*(u)$ of the expansion about the instantonic saddle obey the modified equation
\begin{align}
  2u(n+1) \, Z^*_{n+1}(u) = - 2n\, Z^*_n(u) + 2nu\, Z^*_n{}'(u) + u^2 Z^*_n{}''(u) \;.
\end{align}

The fact that the trumpet partition function is annihilated by $\mathbf{F}(u,t)$ has important implications at $g>0$.  In fact, let us consider the gluing formula \eqref{EQ:higher_genus} for $n=1$. Since the dependence on $u$ and $t$ comes exclusively from the single $Z^{\mathrm{trumpet}}(u,b,t)$ inside the integral, we can immediately conclude that the disk partition function is a solution of the flow equation, not just at genus zero, but at any genus $g$:
\begin{align}
  \mathbf{F}(u,t) \, Z_{g,1}(u;t) = 0 \;.
\end{align}

A similar conclusion can be drawn for other topologies, i.e.\ when $n>1$, but it requires a modification in the way we assign boundary conditions.
So far, we considered a specific way of assigning Dirichlet boundary conditions. Specifically, we imposed on each boundary the same value $\phi_b = 1/\sqrt{t}$ for the dilaton field. In principle, nothing prevents us from computing higher topologies by gluing trumpets associated with different values of $t$. The generalization for the gluing formulas presented in Section \ref{SEC:other_topologies} is actually straightforward, and so is the generalization of the result written in terms of the building blocks \eqref{EQ:Z_building_block}. This is effectively a refinement of the results considered so far, since for any given topology, the generalized partition function $Z_{g,n}$ is now a function of $n$ different deformation parameters $t_1,\ldots,t_n$. At the level of the flow equation, to each boundary is associated a differential operator $\mathbf{F}(u_i,t_i)$ which annihilates the partition function:
\begin{align}
  \mathbf{F}(u_i,t_i) \, Z_{g,n}(u_1,\ldots,u_n;t_1,\ldots,t_n) = 0 \;.
\end{align}
For the purpose of the present paper, we will not consider further this refinement, and we will only deal with the case where the dilaton field takes the same value on each disconnected component of the boundary.

\section{Topological recursion}\label{SEC:topological_recursion}
In Section~\ref{SEC:other_topologies}, we have given a prescription to obtain the resolvents $R_{g,n}$ for any topology. At $t=0$, these functions have a natural interpretation in terms of correlators computed in a certain dual double-scaled Hermitian matrix integral \cite{Saad:2019lba}. The correlators enjoy Schwinger--Dyson-like identities, known as \emph{loop equations}, that allow to recursively compute results at all orders in the large-$N$ expansion \cite{Eynard:2004mh}.
This procedure goes under the name of \emph{topological recursion}, and the set of data initiating it can be captured by a single mathematical object: the \emph{spectral curve} \cite{Eynard:2007kz}.

Thanks to the simplicity of the undeformed trumpet partition function \eqref{EQ:zero_t_limit_Z_trumpet}, at $t=0$ the resolvents $R_{g,n}$ are, essentially, the Laplace transforms of the Weil--Petersson volumes $V_{g,n}$. In fact, the undeformed topological recursion \cite{Saad:2019lba} is given precisely by the recursion formula of Eynard and Orantin \cite{Eynard:2007fi}, which is the Laplace-transformed version of the recursion formula derived by Mirzakhani \cite{Mirzakhani:2006fta}.

Remarkably, we find that the deformation induced by $t$ represents a consistent deformation of the spectral curve.
By this, we mean that the resolvents $R_{g,n}$ computed through the Weil--Petersson gluing, as described in Section~\ref{SEC:other_topologies}, can be reproduced by the topological recursion associated with a $t$-deformation of the Eynard--Orantin spectral curve capturing the $t=0$ case.

To define the spectral curve we introduce the map
\begin{align}
  E(z)
  &= -z^2 \;,
\end{align}
which, in turn, determines the functions
\begin{align}\label{EQ:W01_definition}
  W_{0,1}(z_1;t)
  &= \mathrm{i}\pi \rho(E(z_1),t) \, E'(z_1) \cr
  &= \frac{z_1(2+tz_1^2)}{4\pi} \, \sin\!\Big(\pi z_1{\textstyle\sqrt{4+tz_1^2}}\Big) \;,
\end{align}
and
\begin{align}\label{EQ:W02_definition}
  W_{0,2}(z_1,z_2;t)
  &= \left(R_{0,2}(E(z_1),E(z_2);t) - \frac{1}{(E(z_1)-E(z_2))^2}\right) E'(z_1) \, E'(z_2) \cr
  &= \frac{4(2+tz_1^2)\,(2+tz_2^2)}{\vphantom{\sqrt{z_1^2}}(z_1^2-z_2^2)^2_{\vphantom{1}}\,[4+t(z_1^2+z_2^2)]^2_{\vphantom{1}}} \left(2z_1z_2 + \frac{4(z_1^2+z_2^2) + t(z_1^4+z_2^4)}{\sqrt{4+tz_1^2}\sqrt{2+tz_2^2}}\right) .
\end{align}
Notice how both functions are meromorphic in some neighborhood of the origin.

For any choice of $g$ and $n$ other than the two cases above, the topological recursion computes the $W_{g,n}$ functions through the recursion formula\footnote{
There are more general formulations of the topological recursion. The formulas in \eqref{EQ:topological_recursion_formula} and \eqref{EQ:K_recursion} are valid for a map $E$ such that $\mathrm{d}E$ vanishes at $z=0$, where its local Galois involution is $\sigma:z\mapsto-z$.
}
\begin{align}\label{EQ:topological_recursion_formula}
  W_{g,n}(z_1,\ldots,z_n;t)
  = \operatorname*{Res}_{z\to0} K(z_1,z;t) \, \bigg[ &\,W_{g-1,n+1}(z,-z,z_2,\ldots,z_n;t) \cr
  &+\sum_{\substack{h_1+h_2 = g\\I_1 \cup I_2 = J}}^* W_{h_1,1+|I_1|}(z,I_1;t) \, W_{h_2,1+|I_2|}(-z,I_2;t) \bigg] \;, \cr
\end{align}
where $J = \{z_2, \ldots, z_n\}$, and the symbol $*$ over the sum indicates that one should exclude terms where $(h_1,I_1) = (g,J)$ or $(h_2,I_2) = (g,J)$. The recursion kernel $K$ that appears in \eqref{EQ:topological_recursion_formula} is defined as
\begin{align}\label{EQ:K_recursion}
  K(z_1,z;t)
  &= \frac{1}{2[W_{0,1}(z;t)+W_{0,1}(-z;t)]} \int_{-z}^z \mathrm{d}z_2 \; W_{0,2}(z_1,z_2;t) \cr
  &= \frac{(2+tz_1^2)\sqrt{4+tz_{\phantom{1}}^2}}{(2+tz_{\phantom{1}}^2)\sqrt{4+tz_1^2}} \, \frac{4\pi \csc(\pi z\sqrt{4+tz_{\phantom{1}}^2})}{(z_1^2-z_{\vphantom{1}}^2)\,[4+t(z_1^2+z_{\vphantom{1}}^2)]\vphantom{\sqrt{z_1^2}}} \;.
\end{align}

The functions $W_{g,n}$ computed by the recursion are closely related to the resolvents $R_{g,n}$. Specifically, one can obtain the former by simply acting on the latter with the change of variables induced by $E(z)$,
\begin{align}\label{EQ:W_definition}
  W_{g,n}(z_1,\ldots,z_n;t)
  &= R_{g,n}(E(z_1),\ldots, E(z_n);t) \; E'(z_1) \ldots E'(z_n) \;.
\end{align}
As such, $W_{g,n}$ can be computed, from a bulk perspective, through the gluing \eqref{EQ:higher_genus}. In the spirit of Section~\ref{SEC:other_topologies}, we define
\begin{align}
  \tilde{W}_{m}(z,t)
  &= \tilde{R}_m(E,t) \, E'(z) \cr
  &= (2m+1)! \, \frac{2+tz^2}{(4+tz^2)^{m+3/2}} \left(\frac{2}{z}\right)^{\!2m+2} \;,
\end{align}
which provides the contribution associated with a single trumpet integrated against $b^{2m}$.

To prove that the topological recursion \eqref{EQ:topological_recursion_formula} indeed matches the results obtained in Section~\ref{SEC:other_topologies}, we will show that the $W_{g,n}(z_1,\ldots,z_n;t)$ are connected to the undeformed ones, $W_{g,n}(z_1,\ldots,z_n;0)$, through a change of variables.\footnote{
As it is clear from the definitions in \eqref{EQ:W01_definition}, \eqref{EQ:W02_definition} and \eqref{EQ:W_definition}, the functions $W_{g,n}$ transform as a differential $n$-forms. In fact, the spectral curve and the topological recursion are naturally formulated in terms of differential forms
\begin{align}
  \omega_{g,n}(z_1,\ldots,z_n) = W_{g,n}(z_1,\ldots,z_n) \; \mathrm{d}z_1\wedge\ldots\wedge\mathrm{d}z_n \;.
\end{align}
The recursion kernel $K$, on the other hand, defines a tensor of type $(1,1)$.
} A simple way to show this is to consider an alternative choice for the map $E$. In particular, we consider a map with an explicit dependence on $t$,
\begin{align}\label{EQ:deformed_E_map}
  \hat{E}(\zeta) = -\frac{2}{t}\left(\sqrt{1+t\zeta^2}-1\right) \;,
\end{align}
This choice has two important properties. The first is that it correctly reproduces the undeformed map when $t$ vanishes, i.e. $ \lim_{t\to0} \hat{E}(\zeta) = E(\zeta)$.
The second is that it eliminates any dependence from $t$ in $W_{0,1}$ and $W_{0,2}$, which then necessarily agree with the undeformed ones.
Specifically,
\begin{align}\label{EQ:W01_W02_undeformed}
  \hat{W}_{0,1}(\zeta_1) &= \frac{\zeta_1\sin(2\pi\zeta_1)}{2\pi} & \hat{W}_{0,2}(\zeta_1,\zeta_2) &= \frac{1}{(\zeta_1-\zeta_2)^2} \cr
  &= W_{0,1}(\zeta_1;0) \;, & &= W_{0,2}(\zeta_1,\zeta_2;0) \;.
\end{align}
The same, then, holds true for the recursion kernel
\begin{align}\label{EQ:K_undeformed}
  \hat{K}(\zeta_1,\zeta)
  &= \frac{\pi\csc(2\pi\zeta)}{\zeta_1^2-\zeta_{\vphantom{1}}^2} \cr
  &= K(\zeta_1,\zeta;0) \;.
\end{align}
As anticipated at the beginning of this section, the undeformed recursion induced by \eqref{EQ:W01_W02_undeformed} and \eqref{EQ:K_undeformed} is the Eynard--Orantin topological recursion, while the undeformed $W_{g,n}$ are connected to the Weil--Petersson volumes $V_{g,n}$ by a simple integral transform, as it can be easily seen from the expression of the integrated trumpet
\begin{align}
  \hat{\tilde{W}}_{m}(z)
  &= (2m+1)!\,z^{-2m-2} \cr
  &= \tilde{W}_{m}(z,0) \;.
\end{align}
We can therefore argue as follows. We start from the undeformed case, where the recursion formula is known to hold, and we notice that all quantities can be lifted to the case of finite $t$ through the change of variables induced by $E\circ\hat{E}^{-1}$. Because the recursion formula \eqref{EQ:topological_recursion_formula} is covariant under change of variables induced by maps that are bi-holomorphic in some neighborhood of $z=0$, the topological recursion is guaranteed to hold at any finite $t$.

As an example, we notice that both the recursion and the standard gluing procedure give, for the disk at genus one,
\begin{align}
  W_{1,1}(z_1;t)
  &= \frac{(2+tz_1^2) \, [6+\pi^2z_1^2\,(4+tz_1^2)]}{3 z_1^4\,(4+tz_1^2)^{5/2}} \;.
\end{align}

\section{The spectral form factor}\label{SEC:spectral_form_factor}
At $t=0$ \cite{Saad:2019lba}, JT gravity was observed to reproduce the characteristic shape of a spectral form factor associated with an ensemble of Hamiltonians with random-matrix statistics. From a bulk perspective, the spectral form factor can be interpreted as a transition amplitude in the Hilbert space of two copies of JT gravity \cite{Saad:2019pqd}.
It is computed by the analytic continuation of two boundaries, $u_1\mapsto u+\mathrm{i}\tau$, $u_2\mapsto u-\mathrm{i}\tau$, which introduces a timescale $\tau$. The quantity includes terms coming from different topologies, each weighted by the usual topological factor,
\begin{align}
  F(u,\tau;t)
  &= e^{2S_0}\,Z_{0,1}(u+\mathrm{i}\tau;t)\,Z_{0,1}(u-\mathrm{i}\tau;t) + Z_{0,2}(u+\mathrm{i}\tau,u-\mathrm{i}\tau;t) + \ldots \;,
\end{align}
where the dots correspond to subleading terms associated with higher-genus topologies. We can rewrite the definition in a graphical way as
\begin{align}
  F &= e^{2S_0} \quad \raisebox{-0.45\height}{\sffdisk} \quad + \quad \raisebox{-0.45\height}{\sffcylinder} \quad + \qquad \ldots \;.
\end{align}

Interestingly, different features of the spectral form factor are associated with contributions coming from different topologies.
The initial ``slope'' region comes from considering two disjoint disks \eqref{EQ:Z_disk}. The characteristic shape of the slope can be observed by looking at its large-$\tau$ regime,
\begin{align}\label{EQ:sff_disk}
  Z_{0,1}(u+\mathrm{i}\tau;t)\,Z_{0,1}(u-\mathrm{i}\tau;t) \sim \frac{1}{2\pi\tau^3} \big(1-e^{-4u/t}\big) + \frac{e^{-2u/t}}{\pi\tau^3} \, \sin\!\Big(\frac{2\tau}{t}\Big) \;.
\end{align}
\begin{figure}[tbp]
  \centering
  \includegraphics[]{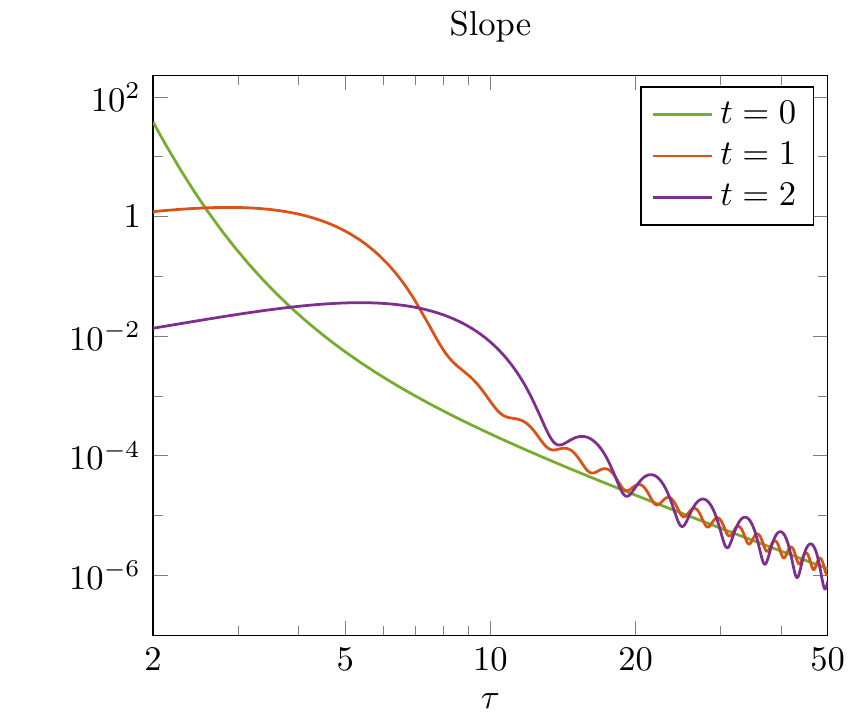}
  \caption{\label{FIG:slope_plot} A log--log plot of the slope $Z_{0,1}(1+\mathrm{i}\tau;t)\,Z_{0,1}(1-\mathrm{i}\tau;t)$ for various values of $t$.}
\end{figure}
The first term gives a cubic decay that reproduces the known $t\to 0$ limit, while the second term is an oscillation of period $\pi t$ whose amplitude is exponentially suppressed in $1/t$.

Eventually, the slope phase will end, and other topologies will dominate the form factor.
The characteristic ``ramp'' region comes, in fact, from the connected topology, i.e., from the cylinder \eqref{EQ:cylinder}, which represents a Euclidean wormhole connecting the two boundaries. The large-$\tau$ behavior is again dominated by two terms,
\begin{align}\label{EQ:sff_cylinder}
  Z_{0,2}(u+\mathrm{i}\tau,u-\mathrm{i}\tau;t) \sim \frac{\tau}{4\pi u}\big(1-e^{-4u/t}\big) - \frac{e^{-2u/t}}{2\pi} \, \cos\!\Big(\frac{2\tau}{t}\Big) \;.
\end{align}
\begin{figure}[tbp]
  \centering
  \includegraphics[]{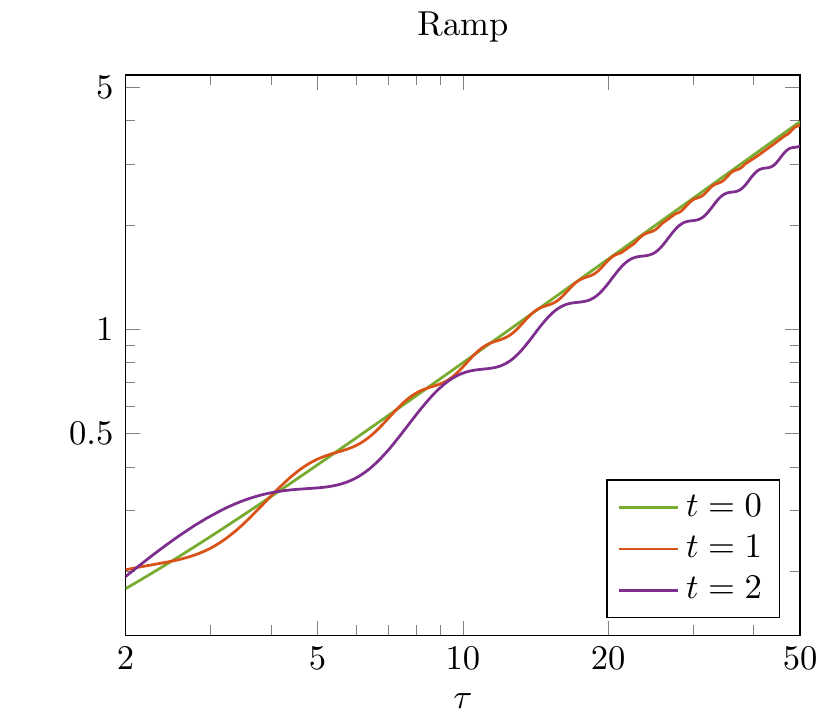}
  \caption{\label{FIG:ramp_plot} A log--log plot of the ramp $Z_{0,2}(1+\mathrm{i}\tau,1-\mathrm{i}\tau;t)$ for various values of $t$.}
\end{figure}
In Figure~\ref{FIG:slope_plot} and Figure~\ref{FIG:ramp_plot}, the slope and the ramp are plotted as functions of $\tau$ for various values of $t$.

In \cite{Saad:2019pqd} the spectral form factor has been discussed using the long-time behavior of the Hartle--Hawking wavefunction associated with JT gravity. The physical origin of the slope, and in particular, of its decaying character, has been interpreted probabilistically. As the time $\tau$ increases, the amplitude for the time evolved Hartle--Hawking state to have a small Einstein--Rosen bridge decreases, being mostly supported at large values of the bridge length. The relevant initial state is localized at small Einstein--Rosen bridge lengths, and thus the transition amplitude decreases with $\tau$. The exchange of a baby universe instead explains the ramp in the spectral form factor: at late times $\tau$, Euclidean wormholes allow transitions from the initial Hartle--Hawking state to a state with a short Einstein--Rosen bridge and a large baby universe, with a size of order $\tau$. The amplitude for this process, while exponentially small in the entropy, does not decay as $\tau$ increases, and the linear growth comes from the $\tau$ different ways in which the baby universe can be rotated before being absorbed \cite{Saad:2019pqd}.

At finite cutoff, we see from \eqref{EQ:sff_disk} and \eqref{EQ:sff_cylinder} that with respect to the undeformed case, the leading $\tau$-behaviours are diminished by exponential finite-size effects while novel periodic fluctuations appear. We observe that both the ramp term \eqref{EQ:sff_cylinder} and the slope \eqref{EQ:sff_disk} share the same damping factor in the non-oscillating part, which indeed suggests we are capturing a universal effect of gravity at finite volume and confirms our result for the cylinder partition function. As a matter of fact, the intersection of these two regimes is independent of the cutoff parameter $t$, up to exponentially-suppressed terms, leading to a transition time of order\footnote{The numerical factors are due to the choice of our conventions but can be simply reabsorbed into a redefinition of the entropy $S_{0}$.}
\begin{equation}
\tau \sim (2u)^{1/4} e^{S_0/2} \;.
\end{equation}
\begin{figure}[tbp]
  \centering
  \includegraphics[]{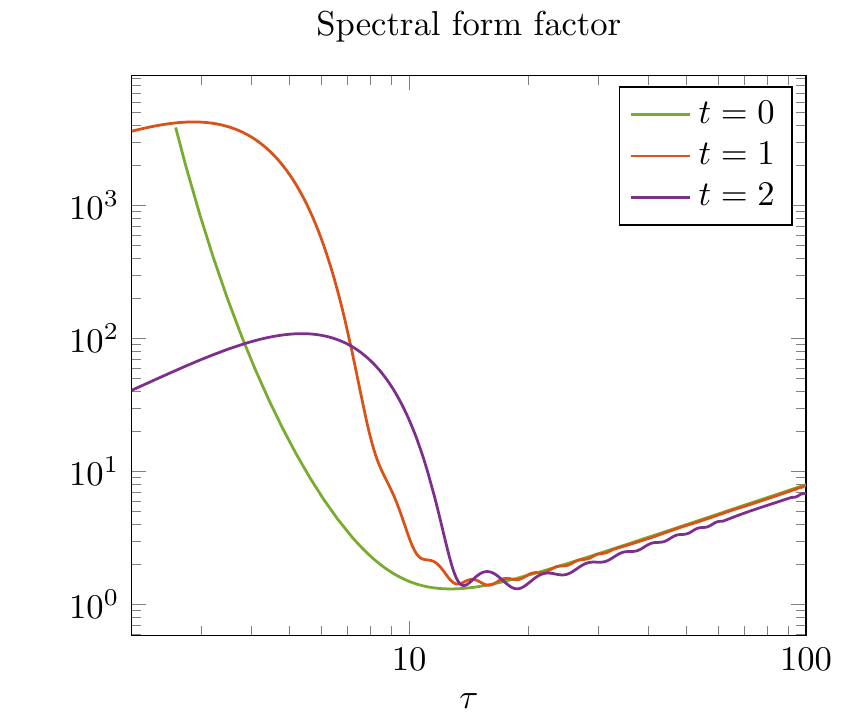}
  \caption{\label{FIG:combo_plot} A log--log plot of the spectral form factor $F(1,\tau,t)$ with $S_0=4$ for various values of $t$. It includes contributions from the slope and the ramp; other topologies are discarded.}
\end{figure}
In Figure~\ref{FIG:combo_plot}, the spectral form factor is plotted as a function of $\tau$ for various values of $t$.

A more physical understanding of the late-time behavior at finite cutoff should come from repeating the analysis of \cite{Saad:2019pqd} using the real Hartle--Hawking wave function obtained in \cite{Iliesiu:2020zld}. It considers both an expanding and a contracting branch, and it is nonperturbative in the parameter $t$: we expect, in particular, an interpretation of the oscillating terms as coming from some interference effects. 

More generally, the appearance of nonperturbative contributions in the disk partition functions seems to suggest the presence of some kind of tunneling process. At $t=0$, the boundary of the AdS$_2$ patch extends to asymptotic infinity, which corresponds to having an infinite potential wall. However, as we move the boundary towards the interior of AdS$_2$ by increasing $t$, the potential barrier turns out to be not infinite anymore, and a hard wall replaces it through Dirichlet conditions. Equivalently, according to the holographic RG picture presented in \cite{McGough:2016lol}, one would expect that integrating out the geometry between the asymptotic boundary and a finite radial distance would result in the $T\bar{T}$ deformation of the original Schwarzian. It is not unreasonable that the effective dynamics could support nontrivial tunneling amplitudes among different vacua in both cases.

\section{Conclusions and outlook}\label{SEC:conclusions}
This paper studied JT gravity as a model for holography at finite volume, considering the nonperturbative contributions coming from the cutoff scale. We assumed the conjectured holographic duality between a $T\bar{T}$-deformed theory and gravity in a finite patch of AdS space. In the case of JT gravity, this amounts to consider the Schwarzian theory deformed by a one-dimensional analog of the $T\bar{T}$ deformation. The same approach has been advocated in \cite{Iliesiu:2020zld} where, among other things, crucial nonperturbative aspects arising from the radial cutoff have been explored. We investigated the appearance of exponentially-suppressed terms in the disk and trumpet partition functions through Borel resummation and resurgence, finding a nice relation between these terms and the analytic structure of the perturbative series in the deformation parameter. We then applied our findings to compute results for arbitrary topologies, exploiting the gluing procedure of \cite{Saad:2019lba}. The construction results in a consistent deformation of the Eynard--Orantin topological recursion relations, although we have not attempted to give a physical interpretation to the deformed spectral density.

There are a certain number of open questions arising from our studies that could stimulate further investigations. In our opinion, a crucial one concerns the physical realization of the deformed holographic dual. As already stressed, the deformed spectral density associated with the topological recursion is not positive definite, and it certainly cannot originate from an ordinary (double-scaled) matrix model. It would be interesting to explore the resolution of this problem in a wider perspective: for example, na\"ive non-positive-definite spectral densities show up in super-matrix models. Actually, in the context of JT gravity, one could consider the supersymmetric version of the bulk theory \cite{Chamseddine:1991fg,Forste:2017kwy}. In this case, the holographic dual is a supersymmetric Schwarzian quantum mechanics: its disk partition function provides nevertheless a positive spectral density that allows considering a higher-genus completion, whose interpretation is given in terms of different matrix ensembles \cite{Stanford:2019vob,Witten:2020bvl}. The $T\bar{T}$ deformation (or some kind of) could arise from a radial cutoff in the bulk theory even for the supersymmetric JT gravity: we expect that exploring this direction would undoubtedly improve the understanding of the present construction and maybe provide a link with the matrix model approach proposed in \cite{Rosso:2020wir}. A somewhat similar positivity problem appears in the context of JT gravity with defects \cite{Maxfield:2020ale} and was solved by an appropriate sum over quantum configurations, leading to a nontrivial modification of the theory \cite{Johnson:2020lns}.

Another research direction worth to be taken into account concerns a full first-principles derivation of the path integral for the bulk theory at finite cutoff. Besides providing a solid foundation for the gluing procedure used in Section~\ref{SEC:other_topologies}, this would unambiguously fix the integration measure, which in \cite{Iliesiu:2020zld} is assumed to be unaltered by the deformation. Such a choice is in sharp contrast with the approach of \cite{Stanford:2020qhm}, where instead, a significant role was played by a particular class of paths in the nonperturbative regime, leading to very different results. In order to gain a better understanding of this discrepancy, one would perform a bulk path-integral calculation, taking properly into account configurations with non-constant Schwarzian action \cite{Moitra:2021uiv}: in principle, a semiclassical calculation could elucidate the contribution of such configurations to the path-integral measure and indicate other nonperturbative effects.

It would also be interesting to understand the behavior of the bulk theory at finite cutoff in the presence of matter: for example, the existence of a $\mathrm{U}(1)$ chiral current could provide a deformation of the Schwarzian quantum mechanics analogous to the $J\bar{T}$ deformation in two-dimensional CFTs \cite{Guica:2017lia}. In \cite{Chakraborty:2020xwo}, such a deformation has been proposed leading, for some choice of the parameters, to a positive-definite spectral density. 

As a final remark, we observe that the spectral form factor in the $T\bar{T}$-deformed theory shows some differences with respect to the original JT case, although certain universal aspects remain present. In particular, the appearance of an oscillating term in the ramp regime hints towards a different interpretation of the holographic picture. 

\section*{Acknowledgments}
We thank Marisa Bonini for interesting discussions. The work of R.P. is supported by the Knut and Alice Wallenberg Foundation under grant Dnr KAW 2015.0083.

\appendix

\section{Perturbative expansion from Bell polynomials}\label{APP:power_expansion}
In this appendix, we show how to obtain the coefficients $A_n$ defined in \eqref{EQ:A_n_coefficients_decomposition}. These determine the $t$-expansion of the disk and the trumpet partition functions. As in Section~\ref{SEC:perturbative_expansion}, we begin by decomposing the exponential term as
\begin{align}
  e^{-I(u,t;s)} = e^{-us^2/2} \, \exp\!\left(us^2\left(\frac12-\frac{1}{\sqrt{1-t s^2}+1}\right)\right) \;,
\end{align}
and define $A_n$ as the coefficients in the expansion of the second term about $t=0$,
\begin{align}\label{EQ:A_tilde_n_coefficients}
  \exp\!\left(us^2\left(\frac12-\frac{1}{\sqrt{1-t s^2}+1}\right)\right)
  &= \sum_{n=0}^{\infty} A_{n} \, t^{n} \;,
\end{align}
where
\begin{align}\label{EQ:A_n_as_derivative}
  A_{n} = \frac{1}{n!} \, \frac{\mathrm{d}^n}{\mathrm{d}t^n} \left.\exp\!\left(us^2\left(\frac12-\frac{1}{\sqrt{1-t s^2}+1}\right)\right)\right|_{t=0} \;.
\end{align}
To compute the $n$-th derivative above, we interpret the exponential as a composite function and make repeated use of the Faà di Bruno formula
\begin{align}\label{EQ:faa_di_bruno}
  \frac{\mathrm{d}^n}{\mathrm{d}t^n}(f \circ g)(t) = \sum_{k=0}^n f^{(k)}(g(t)) \; \mathrm{B}_{n,k}(g'(t),g''(t),\dots,g^{(n-k+1)}(t))
\end{align}
involving Bell polynomials $\mathrm{B}_{n,k}$.

We apply \eqref{EQ:faa_di_bruno} a first time, with the square root in \eqref{EQ:A_n_as_derivative} playing the role of the function $g$. This gives
\begin{align}
  A_{n}
  &= \frac{1}{n!} \sum_{k=0}^{n} C_k \; \mathrm{B}_{n,k}\bigg(\!\left\{-\frac{(2j-3)!!\;s^{2j}}{2^j(1-ts^2)^{j-1/2}}\right\}_{\!j=1,\ldots,n-k+1}\bigg) \bigg|_{t=0} \;,
\end{align}
where
\begin{align}
  C_k = \frac{\mathrm{d}^{k}}{\mathrm{d}z^k} \, \exp\bigg(us^2\bigg(\frac{1}{2}-\frac{1}{z+1}\bigg)\bigg)\bigg|_{z=1} \;.
\end{align}
We then apply \eqref{EQ:faa_di_bruno} again to determine $C_k$. This time, the function $g$ is identified with the term within parenthesis inside the exponential,
\begin{align}
  C_k 
  &= \sum_{l=0}^{k} \frac{\mathrm{d}^{l}}{\mathrm{d}y^l} \, e^{us^2y} \, \bigg|_{y=0} \; \mathrm{B}_{k,l}\bigg(\!\left\{\frac{(-1)^{j+1}j!}{(z+1)^{j+1}}\right\}_{j=1,\ldots,k-l+1}\bigg)\bigg|_{z=1} \;.
\end{align}

Bell polynomials enjoy the identity
\begin{align}\label{EQ:bell_polynomial_identity}
  \mathrm{B}_{n,k}(ab x_1,ab^2 x_2,\dots,ab^{n-k+1}x_{n-k+1}) = a^k b^n \, \mathrm{B}_{n,k}(x_1,x_2,\dots,x_{n-k+1}) \;.
\end{align}
We can make use of the above to rewrite the expressions for $A_n$ and $C_k$ in terms of known combinations of Bell polynomials,
\begin{align}
  A_{n}
  &= \frac{1}{n!} \left(\frac{s^2}{2}\right)^{\!n} \sum_{k=0}^{n} \; (-1)^k \; C_k \; \mathrm{B}_{n,k} \left((-1)!!,1!!,3!!\dots,\left(2(n-k)-1\right)!!\right) \;, \\
  C_k
  &= \frac{1}{(-2)^k} \sum_{l=0}^{k} \left(-\frac{us^2}{2}\right)^l \; \mathrm{B}_{k,l} \left(1!,2!,\dots,\left(k-l+1\right)!\right) \;.
\end{align}
Specifically,
\begin{align}
  \mathrm{B}_{n,k}(1!,2!,3!,(n-k+1)!) &= \binom{n-1}{k-1}\frac{n!}{k!}
\end{align}
are \emph{Lah numbers}, while \cite{qi2017several}
\begin{align}
  \mathrm{B}_{n,k}((-1)!!,1!!,3!!,(2(n-k)-1)!!) &= [2(n-k)-1]!! \; \binom{2n-k-1}{2(n-k)} \;.
\end{align}

By plugging these identities in the expressions above, we arrive at
\begin{align}
  A_n
  &= \frac{1}{2^{2n}n!} \sum_{k=1}^n \frac{(2n-k-1)!}{(n-k)!} \sum_{l=1}^k \binom{k}{l} \bigg({-}\frac{u}{2}\bigg)^{\!l} \frac{s^{2(l+n)}}{(l-1)!} \;.
\end{align}
It is convenient to exchange the two sums with
\begin{align}
  A_n
  &= \frac{1}{2^{2n}n!} \sum_{l=1}^n \bigg({-}\frac{u}{2}\bigg)^{\!l} \frac{s^{2(l+n)}}{(l-1)!} \sum_{k=l}^n \frac{(2n-k-1)!}{(n-k)!} \binom{k}{l} \;.
\end{align}
The final expression \eqref{EQ:A_n_coefficients} is obtained by using
\begin{align}
  \sum_{k=l}^n \frac{(2n-k-1)!}{(n-k)!} \binom{k}{l} &= \frac{(2n)! \, (n-1)!}{(n+l)! \, (n-l)!} \;,
\end{align}
and
\begin{align}
  \sum_{l=1}^n \binom{n-1}{l-1} \frac{(-x)^l}{(n+l)!}
  & = -\frac{(n-1)!}{(2n)!} \, x \, L_{n-1}^{n+1}(x) \;.
\end{align}

\section{Directional Laplace transforms of hypergeometric functions}\label{APP:laplace_transform_hypergeometric}
In this appendix, we collect some useful results concerning the directional Laplace transforms of the hypergeometric functions appearing in the $t$-expansion of the disk and the trumpet partition functions. 

We start from the following identity,
\begin{align}
  \int_{0}^\infty \mathrm{d}\zeta \; e^{-\beta\zeta} \, {}_2F_1\bigg(\frac{1}{2}-m,\frac{1}{2}+m;1;-\alpha\zeta\bigg) = \frac{e^{\beta/2\alpha}}{\sqrt{\pi\alpha\beta}} \, K_{m}\bigg(\frac{\beta}{2\alpha}\bigg) \;,
\end{align}
which holds for $\alpha>0$, $\operatorname{Re}\beta>0$, and $m\in\mathbb{Z}$.
Over the integration range, the hypergeometric is evaluated on the negative real axis. When continuing this result to negative values of $\alpha$, one should be careful about the fact that the Gauss hypergeometric function has a branch cut on the positive real axis. Approaching the branch cut from above and below gives the lateral Laplace transforms
\begin{equation}
\begin{aligned}
  \int_0^{e^{\pm\mathrm{i}0}\infty} \!\!\mathrm{d}\zeta \; e^{-\beta\zeta} \, {}_2F_1\!\left(\frac{1}{2}-m,\frac{1}{2}+m;1;\gamma\zeta\right)
  &= \frac{e^{-\beta/2\gamma}}{\sqrt{-\pi\gamma\beta\mp\mathrm{i}0}} \, K_{m}\!\left(-\frac{\beta}{2\gamma}\pm\mathrm{i}0\right) \cr
  &= \frac{e^{-\beta/2\gamma}}{\sqrt{\pi\gamma\beta}} \left[\pi\,I_m\!\left(\frac{\beta}{2\gamma}\right) \pm (-1)^{m}\mathrm{i}\,K_m\!\left(\frac{\beta}{2\gamma}\right) \right] \;, \cr
\end{aligned}
\end{equation}
where $\gamma>0$. The discontinuity in the directional Laplace transform is reflected in the discontinuity of the square root and of the modified Bessel function $K_m$, the both having a branch cut along the negative real axis. In the last steps of the identity above we used
\begin{align}
  K_m(-x\pm\mathrm{i}0) &= (-1)^{m} K_m(x) \mp \mathrm{i}\pi \, I_m(x) \;,
\end{align}
which holds for $m\in\mathbb{Z}$ and $x>0$.

\begin{figure}[htb]
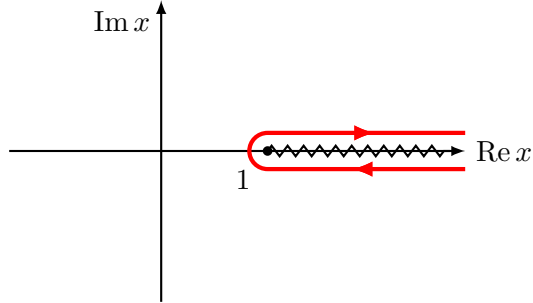

  \centering
  \hankel
  \caption{\label{FIG:Hankel} The Hankel-like contour capturing the discontinuity in the directional Laplace transform, as this crosses the branch cut of the Gauss hypergeometric function at $x\in(1,+\infty)$.}
\end{figure}

The difference between the two lateral Laplace transforms can also be obtained through an integral over a Hankel-like contour wrapping the branch cut, as depicted in Figure~\ref{FIG:Hankel}, which, in turn, amounts to taking the Laplace transform of the discontinuity of the hypergeometric function.
We use the identity
\begin{equation}
\begin{aligned}
  &{}_2F_1(a,b;c;x+\mathrm{i}0) - {}_2F_1(a,b;c;x-\mathrm{i}0) \cr
  &\qquad = \frac{2\pi\mathrm{i}\,\Gamma(c) \, \theta(x-1)}{\Gamma(a)\,\Gamma(b)\,\Gamma(1+c-a-b)} \, (x-1)^{c-a-b} \, {}_2F_1(c-a,c-b;1+c-a-b;1-x) \;, \cr
\end{aligned}
\end{equation}
which holds for $x>1$, to write
\begin{align}
  &\int_0^\infty \mathrm{d}x \; e^{-\beta x} \left[ {}_2F_1\bigg(\frac{1}{2}-m,\frac{1}{2}+m;1;\gamma x+\mathrm{i}0\bigg) - {}_2F_1\bigg(\frac{1}{2}-m,\frac{1}{2}+m;1;\gamma x-\mathrm{i}0\bigg)\right] \cr
  &\qquad = (-1)^m2\mathrm{i} \int_{1/\gamma}^\infty \mathrm{d}x \; e^{-\beta x} \, {}_2F_1\bigg(\frac{1}{2}-m,\frac{1}{2}+m;1;1-\gamma x\bigg) \cr
  &\qquad = (-1)^m2\mathrm{i}e^{-\beta/\gamma} \int_{0}^\infty \mathrm{d}x \; e^{-\beta x} \, {}_2F_1\bigg(\frac{1}{2}-m,\frac{1}{2}+m;1;-\gamma x\bigg) \cr
  &\qquad = (-1)^m2\mathrm{i} \, \frac{e^{-\beta/2\gamma}}{\sqrt{\pi\gamma\beta}} \, K_{m}\bigg(\frac{\beta}{2\gamma}\bigg) \;.
\end{align}

\newpage

\bibliographystyle{JHEP}
\bibliography{bibliography}
\end{document}